\documentclass{intech}
\usepackage[notquote]{hanging}

\booktitle{Will-be-set-by-IN-TECH}%
\chaptertitle{Electroweak Interactions in a Chiral Effective Lagrangian for Nuclei} 
\authors{Brian D. Serot and Xilin Zhang}
\affiliation{Department of Physics and Center for Exploration of Energy and Matter \\
Indiana University--Bloomington, IN 47405}
\country{USA}
\newcommand{\dgamma}[1]{\gamma_{#1}}
\newcommand{\ugamma}[1]{\gamma^{#1}}
\newcommand{\dsigma}[1]{\sigma_{#1}}
\newcommand{\usigma}[1]{\sigma^{#1}}

\newcommand{\dgammafive}{\gamma_{5}}
\newcommand{\ket}[1]{\vert#1\rangle}
\newcommand{\bra}[1]{\langle#1\vert}

\newcommand{\psibar}[1]{\overline{#1}}
\newcommand{\half}{\ensuremath{\frac{1}{2}}}
\newcommand{\threehalf}{\ensuremath{\frac{3}{2}}}
\newcommand{\slashed}[1]{\not\!#1}
\newcommand{\lag}{\mathcal{L}}
\newcommand{\uhalftau}[1]{\frac{\tau^{#1}}{2}}
\newcommand{\dhalftau}[1]{\frac{\tau_{#1}}{2}}
\newcommand{\ucpartial}[1]{\widetilde{\partial}^{#1}}
\newcommand{\dcpartial}[1]{\widetilde{\partial}_{#1}}
\newcommand{\mn}{\mu\nu}
\newcommand{\ab}{\alpha\beta}
\newcommand{\Tdagger}[2]{T^{\dagger \,#1}_{#2}}
\newcommand{\T}[2]{T^{#1}_{\,\,#2}}
\newcommand{\vbg}{\mathsf{v}}
\newcommand{\abg}{\mathsf{a}}
\newcommand{\sbg}{\mathsf{s}}
\newcommand{\pbg}{\mathsf{p}}

\def\Tr{\mathop{\rm Tr}\nolimits}
\begin{document}

\maketitle

\section{Introduction}
\label{sec:intro}
The understanding of electroweak (EW) interactions in nuclei has
played an important role in nuclear and particle physics.
Previously, the electromagnetic (EM) interaction has provided
valuable information about nuclear structure. On the other hand,
weak interactions, which are intrinsically correlated with the EM
interaction, can be complementary to the EM probe. Moreover, a good
knowledge of (anti)neutrino--nucleus scattering cross sections is
needed in other processes, including neutrino-oscillation
experiments, neutrino astrophysics, and others.

To understand EW interactions in nuclei, we need to deal with the
strong interaction that binds nucleons together. The fundamental
theory of the strong interaction is quantum chromodynamics (QCD),
which is a relativistic field theory with local gauge invariance,
whose elementary constituents are colored quarks and gluons. In
principle, QCD should provide a complete description of nuclear
structure and dynamics. Unfortunately, QCD predictions at nuclear
length scales with the precision of existing (and anticipated)
experimental data are not available, and this state of affairs will
probably persist for some time. Even if it becomes possible to use
QCD to describe nuclei directly, this description is likely to be
cumbersome and inefficient, since quarks cluster into hadrons at low
energies.

How can we make progress towards understanding the EW interactions
of nuclei? We will employ a framework based on Lorentz-covariant,
effective quantum field theory and density functional theory.
Effective field theory (EFT) embodies basic principles that are
common to many areas of physics, such as the natural separation of
length scales in the description of physical phenomena. In EFT, the
long-range dynamics is included explicitly, while the short-range
dynamics is parameterized generically; all of the dynamics is
constrained by the symmetries of the interaction. When based on a
local, Lorentz-invariant lagrangian (density), EFT is the most
general way to parameterize observables consistent with the
principles of quantum mechanics, special relativity, unitarity,
gauge invariance, cluster decomposition, microscopic causality, and
the required internal symmetries.

Covariant meson--baryon effective field theories of the nuclear
many-body problem (often called quantum hadrodynamics or QHD) have
been known for many years to provide a realistic description of the
bulk properties of nuclear matter and heavy nuclei. %\citep{Serot86}
[See Refs.~\citep{Serot86, Serot97, Furnstahl03, Serot04}, for
example.] Some time ago, a QHD effective field theory (EFT) was
proposed \citep{Furnstahl9798} that includes all of the relevant
symmetries of the underlying QCD. In particular, the spontaneously
broken $SU(2)_{L} \times SU(2)_{R}$ chiral symmetry is realized
nonlinearly. The motivation for this EFT and illustrations of some
calculated results are discussed in
Refs.~\citep{Furnstahl9798,Huertas02,Huertas03,Huertas04,Twoloops1,Twoloops2,%
Serot07,Threeloops,Serot10}, for example. This QHD EFT has also been
applied to a discussion of the isovector axial-vector current in
nuclei \citep{Ananyan02}.

This QHD EFT has three desirable features: (1) It uses the same
degrees of freedom to describe the currents and the
strong-interaction dynamics; (2) It respects the same internal
symmetries, both discrete and continuous, as the underlying QCD; and
(3) Its parameters can be calibrated using strong-interaction
phenomena, like $\pi$ N scattering and the properties of finite
nuclei (as opposed to EW interactions with nuclei).

In this work, we focus on the introduction of EW interactions in the
QHD EFT, with the Delta (1232) resonance ($\Delta$) included as
manifest degrees of freedom. To realize the symmetries of QCD in QHD
EFT, including both chiral symmetry $SU(2)_{L} \otimes SU(2)_{R}$
and discrete symmetries, we apply the background-field technique
\citep{Gasser84,Serot07}. Based on the EW synthesis in the Standard
Model, a proper substitution of background fields in terms of EW
gauge bosons in the lagrangian, as constrained by the EW
interactions of quarks \citep{Donoghue92}, leads to EW interactions
of hadrons at low energy. This lagrangian has a linear realization
of the $SU(2)_{V}$ isospin symmetry and a nonlinear realization of
the spontaneously broken $SU(2)_{L} \otimes SU(2)_{R}$ (modulo
$SU(2)_{V}$) chiral symmetry (when the pion mass is zero). It was
shown in Ref.~\citep{Furnstahl9798} that by using Georgi's naive
dimensional analysis (NDA) \citep{Georgi93} and the assumption of
naturalness (namely, that all appropriately defined, dimensionless
couplings are of order unity), it is possible to truncate the
lagrangian at terms involving only a few powers of the meson fields
and their derivatives, at least for systems at normal nuclear
densities \citep{Muller96}. It was also shown that a mean-field
approximation to the lagrangian could be interpreted in terms of
density functional theory \citep{Muller96,Serot97,Kohn99}, so that
calibrating the parameters to observed bulk and single-particle
nuclear properties (approximately) incorporates many-body effects
that go beyond Dirac--Hartree theory. Explicit calculations of
closed-shell nuclei provided such a calibration and verified the
naturalness assumption. This approach therefore embodies the three
desirable features needed for a description of electroweak
interactions in the nuclear many-body problem.

Moreover, the technical issues involving spin-3/2 degrees of freedom in
relativistic quantum field theory are also discussed here
\citep{Pascalutsa08,Krebs10}. Following the construction of the
lagrangian, we apply it to calculate certain matrix elements to
illustrate the consequences of chiral symmetries in this theory,
including the conservation of vector current (CVC) and the partial
conservation of axial-vector current (PCAC). To explore the discrete
symmetries, we talk about the manifestation of $G$ parity
in these current matrix elements.

This chapter is organized as follows:  After a short introduction,
we discuss chiral symmetry and discrete symmetries in QCD in the
framework of background fields. The EW interactions of quarks are
also presented, and this indicates the relation between the EW
bosons and background fields. Then we consider the nonlinear
realization of chiral symmetry and other symmetries in QHD EFT, as
well as the EW interactions. Following that, we outline the
lagrangian with the $\Delta$ included. Subtleties concerning the
number of degrees of freedom and redundant interaction terms are
discussed. Finally, some concrete calculations of matrix elements
serve as examples and manifestations of symmetries in the theory. We
also briefly touch on how this formalism can be used to study
neutrino--nucleus scattering
\citep{nunucleon11,inunucleus11,cnunucleus11,Xilinthesis}.

\section{QCD, symmetries, and electroweak synthesis}

In this section, we talk about various symmetries in QCD including
Lorentz-invariance, $C$, $P$, and $T$ symmetries, and approximate
$SU(2)_{L} \otimes SU(2)_{R}$ chiral symmetry (together with baryon
number conservation). The last one is the major focus. Here, we
consider only $u$ and $d$ quarks, and their antiquarks, while others
are chiral singlets. Moreover, the EW interactions, realized in the
electroweak synthesis of the Standard Model, are also discussed with
limited scope.
\subsection{Symmetries}
To consider the symmetries, we apply the background-field technique
\citep{Gasser84}. First we introduce background fields into the QCD
lagrangian, including $\vbg^{\mu}\equiv \vbg^{i\mu} \tau_{i}/{2}$ \
(isovector vector), $\vbg_{(s)}^{\mu}$ \ (isoscalar vector),
$\abg^{\mu}\equiv \abg^{i\mu} \tau_{i}/{2}$ \ (isovector
axial-vector), $\sbg \equiv \sbg^{i} \tau_{i}/{2}$ \ (isovector
scalar), and $\pbg\equiv \pbg^{i} \tau_{i}/{2}$ \ (isovector
pseudoscalar), where $i=x,y,z \ \text{or} \ +1,0,-1$ (the convention
about $i=\pm 1, \ 0 $ will be shown in
Sec.~\ref{subsec:convention}):
\begin{eqnarray}
\lag &=& \lag_{QCD}+ \psibar{q} \dgamma{\mu} (\vbg^{\mu}+ B
       \vbg_{(s)}^{\mu}
        + \dgammafive \abg^{\mu}) q - \psibar{q} (\sbg-i \dgammafive \pbg) q
        \notag \\[5pt]
     &=& \lag_{QCD}+ \psibar{q}_{L} \dgamma{\mu} (l^{\mu}+ B \vbg_{(s)}^{\mu}) q_{L}
        + \psibar{q}_{R} \dgamma{\mu} (r^{\mu}+ B \vbg_{(s)}^{\mu} )q_{R}
        \notag \\[5pt]
     &&\qquad \, \, \, {}-\psibar{q}_{L} (\sbg-i\pbg) q_{R}-
     \psibar{q}_{R} (\sbg+i  \pbg) q_{L} \notag \\[5pt]
     &\equiv& \lag_{QCD}+\lag_{\mathrm{ext}} \ .
     \label{eqn:qcdL}
\end{eqnarray}
Here, $r^{\mu}= \vbg^{\mu}+\abg^{\mu}$,
$l^{\mu}=\vbg^{\mu}-\abg^{\mu}$, $q_{L}=\frac{1}{2}(1-\dgammafive)\,
q$, $q_{R}=\frac{1}{2}(1+\dgammafive)\, q $, $q=(u \ , d)^{T}$ and
$B = 1/3$ is the baryon number. To preserve $C$, $P$, and $T$
invariance of $\lag$, the change of background fields under these
discrete symmetry transformations are determined by the the
properties of the currents coupled to them. The details are in
Tabs.~\ref{tab:bgfieldsundercpt} and \ref{tab:bgfieldsundercpt2}.
Inside the tables,
$\mathcal{P}_{\nu}^{\mu}=\text{diag}(1,-1,-1,-1)_{\mn}$ and
$\mathcal{T}_{\nu}^{\mu}=\text{diag}(-1,1,1,1)_{\mn}$. Moreover, the
Lorentz-invariance is manifest, considering the definition of these
background fields.
\begin{table}
 \centering
   \begin{tabular}{|c|c|c|c|c|c|} \hline
           & $\vphantom{\Big(}\vbg^{\mu}$
           & $\vbg_{(s)}^{\mu}$
           & $\abg^{\mu}$
           & $\sbg$
           & $\pbg$   \\[3pt]  \hline
      $C$  &  $\vphantom{\Big(}-\vbg^{T \mu}$
           & $-\vbg_{(s)}^{\mu}$
           & $\abg^{T \mu}$
           & $\sbg^{T}$
           & $\pbg^{T}$  \\[3pt]   \hline
      $P$  &  $\vphantom{\Big(}\mathcal{P}^{\mu}_{\nu}\vbg^{\nu}$
           &  $\mathcal{P}^{\mu}_{\nu} \vbg_{(s)}^{\nu}$
           &  $-\mathcal{P}^{\mu}_{\nu}\abg^{\nu}$
           &  $\sbg$
           &  $-\pbg$ \\[3pt] \hline
      $T$  & $\vphantom{\Big(}-\mathcal{T}^{\mu}_{\nu}\vbg^{\nu}$
           & $-\mathcal{T}^{\mu}_{\nu} \vbg_{(s)}^{\nu}$
           & $-\mathcal{T}^{\mu}_{\nu} \abg^{\nu}$
           & $\sbg $
           & $-\pbg$  \\[3pt] \hline
   \end{tabular}
   \caption{Transformations of background fields under $C$, $P$, and $T$ operations.
   The transformations of spacetime arguments are not shown here.}
   \label{tab:bgfieldsundercpt}
\end{table}
\begin{table}
 \centering
   \begin{tabular}{|c|c|c|c|c|c|} \hline
           & $\vphantom{\Big(}r^{\mu}$
           & $l^{\mu}$
           & $f_{R\mu\nu}$
           & $f_{L\mu\nu}$
           & $f_{s\mu\nu}$   \\[3pt]  \hline
      $C$  & $\vphantom{\Big(}-l^{T\mu}$
           & $-r^{T\mu}$
           & $-f^{T}_{L\mu\nu}$
           &$-f^{T}_{R\mu\nu}$
           &$-f^{T}_{s\mu\nu}$ \\[3pt]   \hline
      $P$  &  $\vphantom{\Big(}\mathcal{P}^{\mu}_{\nu}l^{\nu}$
           &  $\mathcal{P}^{\mu}_{\nu}r^{\nu}$
           &  $\mathcal{P}_{\mu}^{\lambda}\mathcal{P}_{\nu}^{\sigma}f_{L\lambda\sigma}$
           &  $\mathcal{P}_{\mu}^{\lambda}\mathcal{P}_{\nu}^{\sigma}f_{R\lambda\sigma}$
           &  $\mathcal{P}_{\mu}^{\lambda}\mathcal{P}_{\nu}^{\sigma}f_{s\lambda\sigma}$
           \\[3pt] \hline
      $T$  & $\vphantom{\Big(}-\mathcal{T}^{\mu}_{\nu}r^{\nu}$
           & $-\mathcal{T}^{\mu}_{\nu}l^{\nu}$
           & $-\mathcal{T}_{\mu}^{\lambda}\mathcal{T}_{\nu}^{\sigma} f_{R\lambda\sigma}$
           & $-\mathcal{T}_{\mu}^{\lambda}\mathcal{T}_{\nu}^{\sigma} f_{L\lambda\sigma}$
           & $-\mathcal{T}_{\mu}^{\lambda}\mathcal{T}_{\nu}^{\sigma} f_{s\lambda\sigma}$
           \\[3pt] \hline
   \end{tabular}
   \caption{Continuation of Tab.~\ref{tab:bgfieldsundercpt}.}
   \label{tab:bgfieldsundercpt2}
\end{table}

To understand $SU(2)_L \otimes SU(2)_R \otimes U(1)_B$ symmetry, we can
see that the $\lag$ defined in Eq.~(\ref{eqn:qcdL}) has this symmetry
with the following \emph{local} transformation rules:
\begin{eqnarray}
q_{LA}&\to& \exp \left[-i\frac{\theta(x)}{3} \right]
      \left( \exp \left[ -i\theta_{Li}(x)\,\frac{\tau^{i}}{2} \right] \right)_{A}^{\;B}
      q_{LB} \equiv \exp \left[-i\frac{\theta(x)}{3} \right]
      (L)_{A}^{\;B} q_{LB} \ ,\label{eqn:QCDtransformationrule1} \\[5pt]
q_{R} &\to& \exp \left[ -i\frac{\theta(x)}{3} \right]
      \exp \left[ -i\theta_{Ri}(x)\,\frac{\tau^{i}}{2} \right] q_{R}
      \equiv \exp \left[ -i\frac{\theta(x)}{3} \right] \, R q_{R} \ ,
     \label{eqn:QCDtransformationrule2} \\[5pt]
l^{\mu} &\to& L \, l^{\mu} L^{\dagger} + i L \, \partial^{\mu} L^{\dagger} \ ,
         \label{eqn:QCDtransformationrule3} \\[5pt]
r^{\mu} &\to& R \, r^{\mu} R^{\dagger} + i R \, \partial^{\mu} R^{\dagger} \ ,
            \label{eqn:QCDtransformationrule4} \\[5pt]
\vbg_{(s)}^{\mu} &\to& \vbg_{(s)}^{\mu} - \partial^{\mu} \theta \ ,
 \label{eqn:QCDtransformationrule5} \\[5pt]
\sbg+i\pbg &\to& R(\sbg +i\pbg )L^{\dagger}
       \ .  \label{eqn:QCDtransformationrule6}
\end{eqnarray}

We can also construct field strength tensors that transform
homogeneously:
\begin{eqnarray}
f_{L\mu\nu}&\equiv& \partial_{\mu}l_{\nu}-\partial_{\nu}l_{\mu}
       -i \left[l_{\mu} \, , \, l_{\nu}\right] \to L f_{L\mu\nu}L^{\dagger} \ ,
        \\[5pt]
f_{R\mu\nu}&\equiv& \partial_{\mu}r_{\nu}-\partial_{\nu}r_{\mu}
       -i \left[r_{\mu} \, , \, r_{\nu}\right] \to R f_{R\mu\nu}R^{\dagger} \ ,
       \\[5pt]
f_{s\mu\nu}&\equiv&
       \partial_{\mu}\vbg_{(s)\nu}-\partial_{\nu}\vbg_{(s)\mu} \to
       f_{s\mu\nu}
       \ .
\end{eqnarray}

\subsection{Electroweak synthesis} \label{subsec:EWsyn}
Now we can discuss the electroweak synthesis ($SU_{L}(2)\otimes
U_{Y}(1)$) of the Standard Model, which is mostly summarized in
Tab.~\ref{tab:sm} (electric charge $Q = Y/2 + T_{L}^{3}$)
\citep{IZ80,Donoghue92}. We ignore the Higgs fluctuations and gauge
boson self-interactions:
\begin{table}
  \centering
    \begin{tabular}{|c|c|c|c|c|c|} \hline
          & $T_{L}$  & $T_{L}^{3}$  & $\vphantom{\Big(}Q$    & $Y$
                     & $B$  \\[3pt] \hline
    $u_{L}$ & $\half$  & $\vphantom{\Big(}\half$      & $\frac{2}{3}$ & $\frac{1}{3}$
                     & $\frac{1}{3}$ \\[3pt] \hline
    $d_{L}$ & $\half$  & $\vphantom{\Big(}-\half$     & $-\frac{1}{3}$ & $\frac{1}{3}$
                     & $\frac{1}{3}$ \\[3pt] \hline
    $u_{R}$ &  0       &  0           & $\vphantom{\Big(}\frac{2}{3}$  & $\frac{4}{3}$
                     & $\frac{1}{3}$ \\[3pt] \hline
    $d_{R}$ &  0       &  0           & $\vphantom{\Big(}-\frac{1}{3}$ & $-\frac{2}{3}$
                     & $\frac{1}{3}$ \\[3pt] \hline
    \end{tabular}
    \caption{Multiplets in electroweak synthesis.} \label{tab:sm}
\end{table}
\begin{eqnarray}
\lag_{I}&=&
-\psibar{q_{L}} \ugamma{\mu}(g \frac{\tau_{i}}{2} W^{i}_{\mu}
        +g^{\prime} \frac{Y}{2} B_{\mu})q_{L}
-\psibar{q_{R}}\ugamma{\mu}(g^{\prime}\frac{Y}{2} B_{\mu})q_{R}
        \notag \\[5pt]
        &=& -\psibar{q_{L}} \ugamma{\mu}g (\frac{\tau_{+1}}{2} W^{+1}_{\mu}
        +\frac{\tau_{-1}}{2} W^{-1}_{\mu}) q_{L}
        -\psibar{q_{L}} \ugamma{\mu}(g \frac{\tau_{3}}{2} W^{3}_{\mu}+g^{\prime} \frac{Y}{2} B_{\mu})q_{L} \notag \\[5pt]
           &&{}-\psibar{q_{R}}\ugamma{\mu}(g^{\prime}\frac{Y}{2} B_{\mu})q_{R} \label{eqn:electroweaklag}   \ .
\end{eqnarray}
Here $g$, $g^{\prime}$ and $e$ are the $SU(2)_{L}$, $U(1)_{Y}$ and
$U(1)_{EM}$ charges. To make sure that $U_{EM}(1)$ is preserved, we
impose the following redefinition of excitations relative to the
vacuum ($\theta_{w}$ is the weak mixing angle):
\begin{eqnarray}
B^{\mu}&=&\cos\theta_{w} A^{\mu} -\sin\theta_{w} Z^{\mu} \ , \label{eqn:weakbosonmix1} \\[5pt]
W^{3\mu}&=&\cos\theta_{w} Z^{\mu} +\sin\theta_{w} A^{\mu} \ , \label{eqn:weakbosonmix2} \\[5pt]
g \sin\theta_{w}&=&g^{\prime} \cos\theta_{w}\equiv e
\ . \label{eqn:weakbosonmix3}
\end{eqnarray}

After substituting Eqs.~(\ref{eqn:weakbosonmix1}) to
(\ref{eqn:weakbosonmix3}) into Eq.~(\ref{eqn:electroweaklag}), we
have the right coupling for the EM interaction. Let's compare
Eq.~(\ref{eqn:electroweaklag}) with Eq.~(\ref{eqn:qcdL}); we deduce
the following ($V_{ud}$ describes $u$ and $d$ mixing):
\begin{eqnarray}
l_{\mu}&=&-e\, \frac{\tau^{0}}{2}\, A_{\mu}
       +\frac{g}{\cos\theta_{w}}\sin^{2}\theta_{w}\, \frac{\tau^{0}}{2}\, Z_{\mu} \notag
       \label{eqn.lmubackground} \\[5pt]
&& {}-\frac{g}{\cos\theta_{w}}\frac{\tau^{0}}{2}\, Z_{\mu}
       -g V_{ud}\, \left( W^{+1}_{\mu}\, \frac{\tau_{+1}}{2}
       +W^{-1}_{\mu}\frac{\tau_{-1}}{2} \right) %\notag
       \ , \\[5pt]
r_{\mu}&=&-e\, \frac{\tau^{0}}{2}\, A_{\mu}
       +\frac{g}{\cos\theta_{w}}\sin^{2}\theta_{w}\, \frac{\tau^{0}}{2}\, Z_{\mu} %\notag
       \ , \label{eqn.rmubackground} \\[5pt]
\vbg_{(s)\mu}&=&-e\, \half\,
       A_{\mu}+\frac{g}{\cos\theta_{w}}\sin^{2}\theta_{w}\, \half\,
       Z_{\mu} \ . \label{eqn.vsmubackground}
%\notag
\end{eqnarray}
Furthermore,
\begin{eqnarray}
f_{L\mu\nu}&=&-e\, \uhalftau{0} A_{[\nu , \mu]}
       + \frac{g}{\cos\theta_{w}}\sin^{2}\theta_{w}\, \uhalftau{0}\, Z_{[\nu ,\mu]}
       -\frac{g}{\cos\theta_{w}}\uhalftau{0}\, Z_{[\nu , \mu]} \notag
       \\[5pt]
&& {}-gV_{ud}\, \uhalftau{+1}\, W_{+1[\nu , \mu]}
       -gV_{ud}\, \uhalftau{-1}\, W_{-1[\nu , \mu]}  \notag \\[5pt]
&& {}+ \text{interference \ terms \ including } (WZ) , (WA) , (WW),
       \text{\ but \ no }\
       (ZA) \ , %\notag
       \\[5pt]
f_{R\mu\nu}&=&-e\, \uhalftau{0}\, A_{[\nu , \mu]}
       + \frac{g}{\cos\theta_{w}}\sin^{2}\theta_{w}\, \uhalftau{0}\, Z_{[\nu , \mu]}
       \quad\text{\ (no \ interference \ terms)} \ , %\notag
       \\[5pt]
f_{s\mu\nu}&=& -e\, \half\, A_{[\nu , \mu]}+
       \frac{g}{\cos\theta_{w}}\sin^{2}\theta_{w}\, \half\, Z_{[\nu , \mu]} \ .
       %\notag
\end{eqnarray}
Here
$A_{[\nu , \mu]}\equiv \partial_{\mu}A_{\nu}-\partial_{\nu}A_{\mu}$
and so are the indices of other fields.
If we define [see Eq.~(\ref{eqn:qcdL})]
\begin{eqnarray}
\lag_{\mathrm{ext}}&\equiv&\vbg_{i\mu}V^{i\mu}-\abg_{i\mu}A^{i\mu}
      +\vbg_{(s)\mu}J^{B\mu}
       \notag \\[5pt]
&=& J^{L}_{i\mu}\, l^{i\mu}+J^{R}_{i\mu}\, r^{i\mu}+\vbg_{(s)\mu}J^{B\mu} \ ,
       \label{eqn:currentdefs}
       \\[5pt]
\lag_{I}&=& -eJ^{EM}_{\mu}A^{\mu}
       -\frac{g}{\cos\theta_{w}}J^{NC}_{\mu}Z^{\mu}-gV_{ud}\,
       J^{L}_{+1 \mu}W^{+1\mu}-gV_{ud}\, J^{L}_{ -1 \mu}W^{-1\mu}\ ,
       %\notag
\end{eqnarray}
and use Eqs.~(\ref{eqn.lmubackground}) to
(\ref{eqn.vsmubackground}), we can discover
\begin{eqnarray}
J^{L}_{i \mu} &\equiv& \half\, (V_{i\mu}+A_{i\mu}) \ , %\notag
       \\[5pt]
J^{R}_{i \mu} &\equiv& \half\, (V_{i\mu}-A_{i\mu}) \ , %\notag
       \\[5pt]
J^{EM}_{\mu}&=&V^{0}_{\mu}+\half\, J^{B}_{\mu} \ , %\notag
       \\[5pt]
J^{NC}_{\mu}&=& J^{L0}_{\mu}-\sin^{2}\theta_{w}\, J^{EM}_{\mu} \ .
       \label{eqn:ncdef}
       %\notag
\end{eqnarray}
Here $J^{B}_{\mu}$ is the baryon current, defined to be coupled to
$\vbg_{(s)}^{\mu}$. These relations are consistent with the charge
algebra $Q=T^{0}+B/2$ ($B$ is the baryon number). $V^{i\mu}$ and
$A^{i\mu}$ are the isovector vector current and the  isovector
axial-vector current, respectively. $J^{NC}_{\mu}$, $J^{L}_{\pm1 \mu}$
are the conventional neutral current (NC) and charged current (CC)
up to normalization factors.

\section{QHD EFT, symmetries, and electroweak interactions}
Here we present parallel discussions about QHD EFT's symmetries and
EW interactions. The QHD EFT, as an EFT of QCD at low energy, should
respect all the symmetries of QCD. Moreover, the approximate global
chiral symmetry $SU(2)_L \otimes SU(2)_R \otimes U(1)_B$ in two
flavor QCD is spontaneously broken to $SU(2)_V \otimes U(1)_B$, and
is also manifestly broken due to the small masses of the quarks. To
implement such broken global symmetry in the phenomenological
lagrangian using hadronic degrees of freedom, it was found that
there exists a general nonlinear realization of such symmetry
\citep{Weinb68, Coleman69A, Coleman69B}. Here, we follow the
procedure in Ref.~\citep{Gasser84}. The discussion about the
conventions is presented first.
\subsection{Conventions} \label{subsec:convention}
In this work, the metric $g_{\mn}=\mathrm{diag}(1,-1,-1,-1)_{\mn}$,
and for the Levi--Civita symbol $\epsilon^{\mu\nu\alpha\beta}$, the
convention is $\epsilon^{0123}=1$.
Since we are going to talk about the $\Delta$, which is the lowest
isospin $I=3/2$ nucleon resonance, we define the conventions for
isospin indices. The following example, which shows the relation
between two isospin representations for $\Delta$, may help explain
the convention:
\begin{eqnarray}
\Delta^{\ast a}&\equiv& T^{a}_{\,\,iA} \Delta^{\ast iA} \ . % \notag
\end{eqnarray}
Here $a=\pm 3/2, \pm 1/2$, $i=\pm 1, 0$, and $A=\pm 1/2$. The upper
components labeled as `$a$', `$i$', and `$A$' furnish
$\mathcal{D}^{(3/2)}$, $\mathcal{D}^{(1)}$, and
$\mathcal{D}^{(1/2)}$ representations of the isospin $SU(2)$ group.
(We work with spherical vector components for $I=1$ isospin indices,
which requires some care with signs.) We can immediately realize
that $T^{a}_{\,\,iA}=\langle 1,\frac{1}{2} ;i,A \vert \frac{3}{2};
a\rangle$, which are CG coefficients. It is well known that the
complex conjugate representation of $SU(2)$ is equivalent to the
representation itself, so we introduce a metric linking the two
representations to raise or lower the indices $a,i$, and $A$. For
example, $\Delta_{a}\equiv(\Delta^{\ast a})^{\ast}=T^{\dagger
\,\,iA}_{a} \Delta_{iA}$, where $T^{\dagger \,\,iA}_{a} = \langle
\frac{3}{2}; a \vert 1,\frac{1}{2};i,A \rangle$, should also be able
to be written as
\begin{eqnarray}
\Delta_{a}=T^{iA}_{a} \Delta_{iA} \equiv T^{b}_{jB}\,
\widetilde\delta_{ba}\,\widetilde\delta^{ji}\,\widetilde\delta^{BA}\,\Delta_{iA}
 \ . %\notag
\end{eqnarray}
Here, $\widetilde\delta$ denotes a metric for one of the three
representations. It can be shown that in this convention, $T^{\dagger
\,\,iA}_{a}=T^{iA}_{a}$. Details
about the conventions are given in Appendix~\ref{app:indices}.

\subsection{QHD's symmetry realizations}
Now we proceed to discuss a low-energy lagrangian involving $N^{A}$,
$\Delta^{a}$, $\pi^{i}$, $\rho^{i}_{\mu}$, and the chiral singlets
$V_{\mu}$ and $\phi$ \citep{Furnstahl9798,Serot97}. Under the
transformations shown in Eqs.~(\ref{eqn:QCDtransformationrule1}) to
(\ref{eqn:QCDtransformationrule6}), the symmetry is realized
nonlinearly in terms of hadronic degrees of freedom
\citep{Gasser84}:
\begin{eqnarray}
U&\equiv&\exp \left[ 2i\frac{\pi_{i}(x)}{f_{\pi}}\, t^{i} \right]
       \to  LUR^{\dagger} \ , \label{eqn:allchiral0} \\[5pt]
\xi&\equiv& \sqrt{U}= \exp \left[ i\frac{\pi_{i}}{f_{\pi}}\, t^{i}
       \right]
       \to L\xi h^{\dagger}=h \,\xi R^{\dagger} \ , \label{eqn:allchiral1}  \\[5pt]
\widetilde v_{\mu}&\equiv& \frac{-i}{2} [\xi^{\dagger}(\partial_{\mu}
       -il_{\mu})\xi+\xi(\partial_{\mu}-ir_{\mu})\xi^{\dagger}]
       \equiv \widetilde v_{i\mu}t^{i}
       \to h \, \widetilde v_{\mu} h^{\dagger} -ih \, \partial_{\mu}h^{\dagger} \ ,
       \label{eqn:allchiral2} \\[5pt]
\widetilde a_{\mu}&\equiv& \frac{-i}{2} [\xi^{\dagger}(\partial_{\mu}
       -il_{\mu})\xi-\xi(\partial_{\mu}-ir_{\mu})\xi^{\dagger}]
       \equiv \widetilde a_{i\mu}t^{i} \to h \, \widetilde a_{\mu} h^{\dagger} \ ,
       \label{eqn:allchiral3} \\[5pt]
\dcpartial{\mu}U&\equiv& \partial_{\mu} U -i l_{\mu} U +i U r_{\mu}
       \to L \, \dcpartial{\mu}UR^{\dagger} \ , \label{eqn:allchiral4} \\[5pt]
(\dcpartial{\mu}\psi)_{\alpha}&\equiv& (\partial_{\mu}+i\,\widetilde v_{\mu}
       -i\vbg_{(s)\mu}B)_{\alpha}^{\;\beta} \psi_{\beta}
       \to \exp \left[ -i\theta(x)B \right]
       h_{\alpha}^{\;\beta} (\dcpartial{\mu}\psi)_{\beta} \ , \label{eqn:allchiral5}
       \\[5pt]
\widetilde{v}_{\mu\nu}&\equiv& -i [\widetilde{a}_{\mu} \, , \,
       \widetilde{a}_{\nu}]
       \to h \, \widetilde{v}_{\mu\nu} h^{\dagger} \ ,  \label{eqn:allchiral6}
       \\[5pt]
F^{(+)}_{\mu\nu}&\equiv&\xi^{\dagger}f_{L\mu\nu}\,\xi
       + \xi f_{R\mu\nu}\, \xi^{\dagger}
       \to hF^{(+)}_{\mu\nu}h^{\dagger} \ , \label{eqn:allchiral7} \\[5pt]
F^{(-)}_{\mu\nu}&\equiv&\xi^{\dagger}f_{L\mu\nu}\,\xi
       - \xi f_{R\mu\nu}\, \xi^{\dagger}
       \to hF^{(-)}_{\mu\nu}h^{\dagger} \ , \label{eqn:allchiral8} \\[5pt]
\dcpartial{\lambda} F^{(\pm)}_{\mu\nu} &\equiv&
       \partial_{\lambda}F^{(\pm)}_{\mu\nu} + i [\widetilde{v}_{\lambda}
       \, , \, F^{(\pm)}_{\mu\nu}] \to h\,\dcpartial{\lambda}
F^{(\pm)}_{\mu\nu}h^{\dagger} \ . \label{eqn:allchiral9} %\notag
\end{eqnarray}
In the preceding equations, $t^{i}$ are the generators of reducible
representations of $SU(2)$. Specifically, they could be generators
of $\mathcal{D}^{(1/2)}_{N} \oplus \mathcal{D}^{(1)}_{\rho} \oplus
\mathcal{D}^{(3/2)}_{\Delta}$, which operate on non-Goldstone
isospin multiplets including the nucleon, $\rho$ meson, and
$\Delta$. We generically label these fields by $\psi_{\alpha}=\left(
N_{A}, \rho_{i}, \Delta_{a} \right)_{\alpha}$. Most of the time, the
choice of $t^{i}$ is clear from the context. $B$ is the baryon
number of the particle. The transformations of the isospin and
chiral singlets $V_{\mu}$ and $\phi$ are trivial $(\phi \to \phi$,
$V_{\mu} \to V_{\mu})$. $h$ is generally a local $SU(2)_{V}$ 
matrix. We also make use of the dual field tensors,
for example, $\psibar{F}^{\,(\pm)\,\mu\nu} \equiv
\epsilon^{\mu\nu\alpha\beta} F^{(\pm)}_{\alpha\beta}$, which have
the same chiral transformations as the ordinary field tensors. Here
we do not include the background fields $\sbg$ and $\pbg$ mentioned
in Eq.~(\ref{eqn:qcdL}), which are the source of manifest
chiral-symmetry breaking in the Standard Model.

The $C$, $P$, and $T$ transformation rules are summarized in
Tabs.~\ref{tab:cpteft} and \ref{tab:cpteft2}. A plus sign means
normal, while a minus sign means abnormal, i.e., an extra minus sign
exists in the transformation. The convention for \emph{Dirac
matrices} sandwiched by nucleon and/or $\Delta$ fields are
\begin{eqnarray}
C\psibar{N}\,\Gamma N C^{-1} &=& \begin{cases}
                               -N^{T}\, \Gamma^{T}\,  \psibar{N}^{T}\ ,
                               &\text{normal}\ ; \\
                               N^{T}\, \Gamma^{T} \, \psibar{N}^{T}\ ,
                               &\text{abnormal}\ .
                               \end{cases}  \label{eqn:cconjugdia} %\\ [5pt]
\end{eqnarray}
\begin{eqnarray}
C(\psibar{\Delta}\, \Gamma N + \psibar{N}\, \Gamma \Delta) C^{-1}
&=&
\begin{cases}
        - \Delta^{T}\,\Gamma^{T}\,\psibar{N}^{T}
        -N^{T}\,\Gamma^{T}\,\psibar{\Delta}^{T}\ ,  &\text{normal}\ ; \\
        +\Delta^{T}\,\Gamma^{T}\,\psibar{N}^{T}
        +N^{T}\,\Gamma^{T}\, \psibar{\Delta}^{T}\ ,
        &\text{abnormal}\ .
        \end{cases} \label{eqn:cconjugnondiagA}\\ [5pt]
Ci(\psibar{\Delta}\, \Gamma N - \psibar{N}\, \Gamma \Delta) C^{-1}
&=&
\begin{cases}
        +i \Delta^{T}\,\Gamma^{T}\,\psibar{N}^{T}
        -i N^{T}\,\Gamma^{T}\,\psibar{\Delta}^{T}\ , &\text{normal}\ ; \\
        -i\Delta^{T}\,\Gamma^{T}\,\psibar{N}^{T}
        +i N^{T}\,\Gamma^{T}\,\psibar{\Delta}^{T}\ ,
        &\text{abnormal}\ .
                                      \end{cases} \label{eqn:cconjugnondiagB}
\end{eqnarray}
\noindent Here, in
Eqs.~(\ref{eqn:cconjugdia}),~(\ref{eqn:cconjugnondiagA}),
and~(\ref{eqn:cconjugnondiagB}), the extra minus sign arises because
the fermion fields anticommute. The factor of $i$ in
Eq.~(\ref{eqn:cconjugnondiagB}) is due to the requirement of
Hermiticity of the lagrangian. To make the analysis easier for
$\psibar{\Delta}\,\Gamma N + C.C.$, we can just attribute a minus
sign to an $i$ under the $C$ transformation. Whenever an $i$ exists,
the lagrangian takes the form $i(\psibar{\Delta}\,\Gamma
N-\psibar{N}\,\Gamma \Delta)$. When no $i$ exists, the lagrangian
is like $\psibar{\Delta}\,\Gamma N+\psibar{N}\,\Gamma \Delta$.

For $P$ and $T$ transformations, the conventions are the same for
$N$ and $\Delta$ fields, except for an extra minus sign in the
parity assignment for each $\Delta$ field \citep{Weinb95}, so we list
only the $N$ case:
\begin{eqnarray}
P\psibar{N}\,\Gamma_{\mu} N P^{-1} &=& \begin{cases}
          \psibar{N}\, \mathcal{P}_{\mu}^{\nu}\, \Gamma_{\nu}\, N\ ,
          &\text{normal}\ ; \\
          -\psibar{N}\, \mathcal{P}_{\mu}^{\nu}\, \Gamma_{\nu}\, N\ ,
          &\text{abnormal}\ .
                               \end{cases}  \\ [5pt]
T\psibar{N}\,\Gamma_{\mu} N T^{-1} &=& \begin{cases}
          \psibar{N}\, \mathcal{T}_{\mu}^{\nu}\, \Gamma_{\nu}\, N\ ,
          &\text{normal} \ ; \\
          -\psibar{N}\, \mathcal{T}_{\mu}^{\nu}\, \Gamma_{\nu}\, N\ ,
          &\text{abnormal}\ .
                               \end{cases}
\end{eqnarray}
It is easy to generalize these results to $\Gamma_{\mn}$, etc.

Now a few words about isospin structure are in order.
Suppose an isovector object is denoted as $O_{\mu}\equiv
O_{i\mu}t^{i}$, then the conventions are explained below:
\begin{eqnarray}
CO_{\mu}C^{-1}&=&\begin{cases}
           O_{\mu}^{T} \ ,    &\text{normal} \ ; \\
           -O_{\mu}^{T} \ ,   &\text{abnormal} \ .
    \end{cases} %\notag
    \\[5pt]
PO_{\mu}P^{-1}&=&\begin{cases}
           \mathcal{P}_{\mu}^{\nu}O_{\nu} \ , &\text{normal} \ ; \\
           -\mathcal{P}_{\mu}^{\nu}O_{\nu} \ ,  &\text{abnormal} \ .
    \end{cases}    %\notag
    \\[5pt]
TO_{\mu}T^{-1}&=&\begin{cases}
           \mathcal{T}_{\mu}^{\nu}O_{\nu} \ , &\text{normal} \ ; \\
           -\mathcal{T}_{\mu}^{\nu}O_{\nu} \ , &\text{abnormal} \ .
    \end{cases}   %\notag
\end{eqnarray}
The same convention applies to the isovector (pseudo)tensors. For
isovector (pseudo)scalars, the $\mathcal{P}$ and $\mathcal{T}$
should be changed to $\mathbf{1}$. For the $C$ transformation,
$O^{T}$ means transposing both isospin and Dirac matrices in the
definition of $O$, if necessary.
\begin{table}
  \centering
    \begin{tabular}{|c|c|c|c|c|c|c|c|c|} \hline
     & $\ugamma{\mu}$
     &$\sigma^{\mu\nu}$
     &$1$
     &$\ugamma{\mu}\dgammafive$
     &$i\dgammafive$
     &$i$
     &$i\overset{\leftrightarrow}{\partial}$
     &$\epsilon^{\mu\nu\alpha\beta}$
     \\[2pt] \hline
$C$  & $-$ & $-$ & $+$ & $+$ & $+$ & $-$ & $-$ & $+$  \\[2pt] \hline
$P$  & $+$ & $+$ & $+$ & $-$ & $-$ & $+$ & $+$ & $-$ \\[2pt] \hline
$T$  & $-$ & $-$ & $+$ & $-$ & $-$ & $-$ & $-$ & $-$  \\[2pt] \hline
    \end{tabular}
    \caption{Transformation properties of objects under $C$, $P$, and $T$.
    Here `$+$' means normal and `$-$' means abnormal.}
    \label{tab:cpteft}
\end{table}
\begin{table}
  \centering
    \begin{tabular}{|c|c|c|c|c|c|c|c|c|c|c|c|c|c|} \hline
     &$\widetilde{a}_{\mu}$
     &$\widetilde{v}_{\mu}$
     &$\widetilde{v}_{\mu\nu}$
     &$\rho_{\mu}$
     &$\rho_{\mu\nu}$
     &$\psibar{\rho}_{\mu\nu}$
     &$V_{\mu}$
     &$V_{\mu\nu}$
     &$\psibar{V}_{\mu\nu}$
     &$F^{(\pm)}_{\mu\nu}$
     &$f_{s\mu\nu}$
     &$\psibar{F}^{\,(\pm)}_{\mu\nu}$
     &$\psibar{f}_{s\mu\nu}$  \\[2pt] \hline
$C$   & $+$ & $-$ &
     $-$ & $-$
     & $-$ &$-$ &$-$ &$-$ &$-$ & $\mp$   & $-$ & $\mp$     & $-$  \\[2pt] \hline
$P$   & $-$ & $+$ &
     $+$ & $+$
     & $+$ &$-$ &$+$ &$+$ &$-$ & $\pm$   & $+$ & $\mp$     & $-$  \\[2pt] \hline
$T$  & $-$ & $-$ &
     $-$ & $-$
     & $-$ &$+$ &$-$ &$-$ &$+$ & $-$     & $-$ & $+$       & $+$ \\[2pt] \hline
    \end{tabular}
    \caption{Continuation of Tab.~\ref{tab:cpteft}.}
    \label{tab:cpteft2}
\end{table}

\subsection{QHD EFT lagrangian (without \texorpdfstring{$\Delta$}{Delta}) and electroweak interactions}
Now we begin to discuss the QHD EFT lagrangian. Based on the
symmetry transformation rules discussed above, we can construct the
lagrangian as an invariant of these transformations by using the
building blocks shown in Eqs.~(\ref{eqn:allchiral0}) to
(\ref{eqn:allchiral9}). In principle, there are an infinite number
of possible interaction terms in this lagrangian. However, power
counting \citep{Furnstahl9798,Twoloops1,Twoloops2} and Naive
Dimensional Analysis (NDA) \citep{GEORGI84,Georgi93} enable us to
truncate this series of interactions to achieve a good
approximation. Following the discussion in
Ref.~\citep{Furnstahl9798}, we associate with each interaction term
a power-counting index:
\begin{equation}
\hat{\nu} \equiv d+ \frac{n}{2} + b \ . %\notag
\end{equation}
Here $d$ is the number of derivatives (small momentum transfer) in
the interaction, $n$ is the number of fermion fields, and $b$ is the
number of heavy meson fields.

The QHD theory has been developed for some time. Details can be
found in Refs.~\citep{Furnstahl9798, Serot07, Serot97}. \emph{Here, we give a
complete treatment of electroweak interactions in this theory.} 
(However, we do not discuss ``seagull'' terms of higher order 
in the couplings.) We begin with
\begin{eqnarray}
\lag_{N (\hat{\nu}\, \leqslant\, 3)}&=&
        \psibar{N}(i\ugamma{\mu}[\dcpartial{\mu}
        +ig_{\rho}\rho_{\mu}+ig_{v}V_{\mu}]+g_{A}\ugamma{\mu}\dgammafive\,
        \widetilde{a}_{\mu}-M+g_{s}\phi)N  \notag \\[5pt]
&& {}-\frac{f_{\rho}g_{\rho}}{4M}\, \psibar{N}\rho_{\mu\nu}
        \usigma{\mu\nu}N
        -\frac{f_{v}g_{v}}{4M}\, \psibar{N}V_{\mu\nu} \usigma{\mu\nu}N
        -\frac{\kappa_{\pi}}{M}\, \psibar{N}\,\widetilde{v}_{\mu\nu}
        \usigma{\mu\nu}N   \notag \\[5pt]
&& {}+\frac{4\beta_{\pi}}{M}\, \psibar{N}N \Tr(\widetilde{a}_{\mu}\widetilde{a}^{\mu})
        +\frac{i\kappa_{1}}{2M^{2}}\, \psibar{N}
        \dgamma{\mu}\overset{\leftrightarrow}{\dcpartial{\nu}} N
        \Tr\left(\widetilde{a}^{\mu}\widetilde{a}^{\nu}\right)    \notag
        \\[5pt]
&& {}+\frac{1}{4M}\, \psibar{N} \usigma{\mu\nu}( 2
        \lambda^{(0)}f_{s\mu\nu}+\lambda^{(1)}F^{(+)}_{\mu\nu} ) N \ ,
        \label{eqn:Nlaglowest}
\end{eqnarray}
where $\dcpartial{\mu}$ is defined in Eq.~(\ref{eqn:allchiral5}),
$\overset{\leftrightarrow}{\dcpartial{\nu}} \equiv \dcpartial{\nu} -
(\overset{\leftarrow}{\partial_{\nu}} - i\widetilde v_{\nu}+
i\vbg_{(s)\nu})$, and the new field tensors are $V_{\mu\nu} \equiv
\partial_{\mu} V_{\nu} - \partial_{\nu} V_{\mu}$ and
\begin{equation}
\rho_{\mn} \equiv \partial_{[\mu}\rho_{\nu
        ]}+i\overline{g}_{\rho}[\rho_{\mu}\,
        , \, \rho_{\nu}]
        + i ([\widetilde{v}_{\mu}\, , \, \rho_{\nu}] - \mu \leftrightarrow
        \nu) \to h \, \rho_{\mn} h^{\dagger}\ .
\end{equation}
The superscripts ${}^{(0)}$ and ${}^{(1)}$ denote the isospin. In
Appendix~\ref{app:tildeobjects}, details about the tilde objects
(which are defined exactly above) are shown explicitly in terms of
pion and background fields.

Next is a purely mesonic lagrangian:
\begin{eqnarray}
\lag_{\mathrm{meson} (\hat{\nu}\, \leqslant\, 4)} &=& \half \,
       \partial_{\mu}\phi\,\partial^{\mu}\phi
       + \frac{1}{4} f^{2}_{\pi} \Tr[\dcpartial{\mu}U(\ucpartial{\mu}U)^{\dagger}]
       +\frac{1}{4} f^{2}_{\pi}\, m^{2}_{\pi}\Tr(U+U^{\dagger}-2) \notag
       \\[5pt]
&& {}-\half \Tr(\rho_{\mu\nu}\rho^{\mu\nu}) -\frac{1}{4} \, V^{\mu\nu}V_{\mu\nu}
       \notag \\[5pt]
&& {}+\half \left(1+\eta_{1}\frac{g_{s}\phi}{M}
       +\frac{\eta_{2}}{2}\frac{g^{2}_{s}\phi^{2}}{M^{2}}\right)m^{2}_{v}\, V_{\mu}V^{\mu}
       + \frac{1}{4!}\, \zeta_{0} \, g^{2}_{v}(V_{\mu}V^{\mu})^{2} \notag
       \\[5pt]
&& {}+\left(1+\eta_{\rho}\frac{g_{s}\phi}{M}\right) m^{2}_{\rho} \Tr(\rho_{\mu}\rho^{\mu})
       -\left(\half+\frac{\kappa_{3}}{3!}\frac{g_{s}\phi}{M}
       +\frac{\kappa_{4}}{4!}\frac{g_{s}^{2}\phi^{2}}{M^{2}}\right)m_{s}^{2}\phi^{2}
       \notag \\[5pt]
&& {}+\frac{1}{2g_{\gamma}} \left(
       \Tr(F^{(+)\mu\nu}\rho_{\mu\nu})+\frac{1}{3}\, f_{s}^{\mu\nu}V_{\mu\nu}
       \right) \ .  \label{eqn:lmeson}
\end{eqnarray}
The $\nu=3$ and $\nu=4$ terms in $\lag_{\mathrm{meson} (\hat{\nu}\,
\leqslant\, 4)}$ are important for describing the bulk properties of
nuclear many-body systems \citep{Furnstahl9798,FTS95,FST96}. The
only manifest chiral-symmetry breaking is through the nonzero pion
mass. It is well known that there are other $\hat{\nu}=4$ terms
involving pion-pion interactions. Since multiple pion interactions
and chiral-symmetry-violating terms other than the pion mass term
are not considered, this additional lagrangian is not shown here.

Finally, we have
\begin{eqnarray}
\lag_{N,\pi (\hat{\nu}\, =\, 4)}&=&\frac{1}{2M^{2}}\,
       \psibar{N}\dgamma{\mu}(2\beta^{(0)}
       \partial_{\nu}f_{s}^{\mu\nu}+\beta^{(1)}\dcpartial{\nu}F^{(+)\mu\nu}
       +\beta_{A}^{(1)}\dgammafive \dcpartial{\nu}F^{(-)\mu\nu})N   \notag
       \\[5pt]
&& {}-\omega_{1}\Tr(F^{(+)}_{\mu\nu}\, \widetilde{v}^{\mu\nu})
       +\omega_{2} \Tr(\widetilde{a}_{\mu}\dcpartial{\nu}F^{(-)\mu\nu})
       +\omega_{3} \Tr \left( \widetilde{a}_{\mu} i
       \left[\widetilde{a}_{\nu} \, , \, F^{(+)\mu\nu} \right]\right)  \notag
       \\[5pt]
&& {}-g_{\rho\pi\pi} \, \frac{2f^{2}_{\pi}}{m^{2}_{\rho}}
       \Tr(\rho_{\mu\nu}\widetilde{v}^{\mu\nu})  \notag  \\[5pt]
&& {}+\frac{c_{1}}{M^{2}}\, \psibar{N}\ugamma{\mu} N
       \Tr \left(\widetilde{a}^{\nu}\, \psibar{F}^{(+)}_{\mu\nu}\right)
       +\frac{e_{1}}{M^{2}}\, \psibar{N}\ugamma{\mu}\,
       \widetilde{a}^{\nu}N \,
       \psibar{f}_{s\mu\nu} \notag \\[5pt]
&& {}+\frac{c_{1\rho}g_{\rho}}{M^{2}}\, \psibar{N}\ugamma{\mu} N\Tr
       \left(\widetilde{a}^{\nu}\, \psibar{\rho}_{\mu\nu}\right)
       +\frac{e_{1v}g_{v}}{M^{2}}\, \psibar{N}\ugamma{\mu}\,
       \widetilde{a}^{\nu}N \, \psibar{V}_{\mu\nu}\ . %\notag
\end{eqnarray}
Note that $\lag_{N,\pi (\hat{\nu}\, =\, 4)}$ is \emph{not} a complete list
of all possible $\hat{\nu}=4$ interaction terms. However,
$\beta^{(0)}$ and $\beta^{(1)}$ are used in the form factors of the
nucleon's vector current, $\omega_{1,2,3}$ contribute to the form
factor of the pion's vector current, and $g_{\rho\pi\pi}$ is used in
the form factors that incorporate vector meson dominance (VMD).%
\footnote{VMD in QHD EFT has been discussed in detail in
Ref.~\citep{Serot07}. We will discuss VMD for the form factor of the
transition current involving $\Delta$ and $N$.} Special attention
should be given to the $c_{1}, e_{1}, c_{1\rho}$, and $e_{1v}$
couplings, since they are the only relevant $\hat{\nu}=4$ terms for
NC photon production \citep{inunucleus11,cnunucleus11}.

The construction of these high-order terms, $\lag_{N,\pi (\hat{\nu}
=4)}$ for example, is carried out by exhaustion. Based on the
various symmetry transformation rules, at a given order there are a
finite number of interaction terms, although the number can be big.
For example, the interaction terms involving two pions and only one
nucleon at $\hat{\nu}=4$ \emph{without chiral symmetry breaking} are
\citep{tang98}
\begin{eqnarray}
&& \psibar{N}
\usigma{\mn}i\overset{\leftrightarrow}{\ucpartial{\lambda}} N
\Tr\left(\dcpartial{\lambda}\widetilde{a}_{\mu}
\widetilde{a}_{\nu}\right) \qquad
\text{and \ other \ contractions \ of \ Lorentz \ indices} \notag \ , \\[5pt]
&&\psibar{N} \dgamma{\mu} i \left[ \ucpartial{\mu}
\widetilde{a}^{\nu} \ , \  \widetilde{a}_{\nu}\right]N \qquad
\text{and other \ contractions \ of \ Lorentz \ indices} \ . \notag
\end{eqnarray}

\subsection{Introducing \texorpdfstring{$\Delta$}{Delta} resonances}

The pathologies of relativistic field theory with spin-3/2 particles
have been investigated in the canonical quantization framework for
some time. There are two kinds of problems: one is the so-called
Johnson--Sudarshan problem \citep{JohnsonSundarshan, CRHagen,
Capri1980}; the other one is the Velo--Zwanzinger problem
\citep{VeloZwanzinger, LPSSingh73, Capri1980}. It was realized in
\citep{Takahashi} that the two problems may both be related to the
fact that the classical equation of motion, as the result of
minimizing the action, fails to eliminate redundant spin components,
because the invertibility condition of the constraint equation is
not satisfied all the time. For example, in the Rarita--Schwinger
formalism, the representation of the field is $\psi^{\mu}$: $(\half,
\half) \otimes \left((\half, 0) \oplus (0, \half)\right) =(1, \half)
\oplus (\half, 1) \oplus (\half, 0) \oplus (0, \half)$
\citep{Weinb95fieldreps}. It can be shown that the spin-1/2
components are not dynamical in the free theory, which is generally
not true after introducing interactions. Another issue is about the
so-called \emph{off-shell couplings}, which have the form
$\gamma_{\mu} \psi^{\mu}$, $\partial_{\mu} \psi^{\mu},
\overline{\psi}^{\mkern3mu\mu} \gamma_{\mu},$ and $\partial_{\mu}
\overline{\psi}^{\mkern3mu\mu}$ (still in the Rarita--Schwinger
representation).

Recently, the problem has been investigated in a path-integral
formalism in Ref.~\citep{Pascalutsa98}, where a gauge invariance is
required for interactions. But this constraint conflicts with the
manifest nonlinear chiral-symmetry realization in chiral EFT.
Subsequently, in \citep{Pascalutsa01,Krebs09}, the authors realized
that the commonly used non-invariant interactions are related to
gauge-invariant interactions by field redefinitions, up to some
contact interaction terms. Moreover, from the modern chiral EFT
viewpoint, it has been concluded \citep{tang96,Krebs10} that the
off-shell couplings are redundant, since they lead to contributions
to contact interactions without spin-3/2 degrees of freedom.
Furthermore, it has been proved that off-shell couplings with
$\partial_{\mu}$ changed to $\dcpartial{\mu}$ are also redundant,
which makes the manifest realization of chiral symmetry possible
with a spin-3/2 particle.

However, the modern argument, which makes use of field redefinitions
and gauge invariance for the EFT, looks abstract. The whole argument
is that the field redefinitions, constructed to transform
non-invariant terms to gauge-invariant terms, is applicable here,
which requires us to be far away from the singularities of these
transformations, i.e., to stay at low-energy and in weak-field
regions \citep{Krebs09}. This leads us to give another interesting
argument, based directly on this assumption. In the Hamiltonian
formalism, these two issues are well clarified,\footnote{In the
perturbative calculation of EFT, the time-ordered free propagator
defined in the Hamiltonian formalism for a spin-3/2 particle always
satisfies the constraint on the degrees of freedom. (Assume we have a
well defined Hamiltonian for the EFT.) For finite sums of the series
of diagrams involving this propagator, the constraint is always
satisfied. Moreover, those off-shell terms when either contracted to
external legs or to the internal propagator of spin-3/2 degrees of
freedom, give zero value. We may conclude that they are redundant.
However, it is not clear whether the two conclusions hold for
infinite sums. Moreover, as we know, the time-ordered propagator is
not covariant, and leads to the difficulty of understanding
Lorentz-invariance.} however the quantization of the EFT and hence
Lorentz-invariance are not straightforward. So we use the
path-integral approach.

Let's focus on the spin-3/2 propagator in the Rarita--Schwinger
representation. First, we can decompose the free propagator into
different spin components:
\begin{eqnarray}
S_{F}^{ 0 \mn}(p) 
&=& \frac{-(\slashed{p}+m)}{p^{2}-m^{2}+i\epsilon} \left[g^{\mn}-\frac{1}{3}\ugamma{\mu}\ugamma{\nu}+\frac{p^{\mu}\ugamma{\nu}-p^{\nu}\ugamma{\mu}}{3m}-\frac{2}{3m^{2}}p^{\mu}p^{\nu} \right] \notag \\
&\equiv& -\frac{1}{\slashed{p}-m+i \epsilon}
P^{(\frac{3}{2})\mn} -\frac{1}{\sqrt{3}m}P^{(\half) \mn}_{12}
-\frac{1}{\sqrt{3}m}P^{(\half) \mn}_{21} \notag \\[5pt]
&&+\frac{2}{3m^{2}}(\slashed{p}+m) P^{(\half) \mn}_{22} \ ,
\label{eq:Scomponent1}
%\notag
\\[5pt]
P^{(\frac{3}{2}) \mn}
&=& g^{\mn}-\frac{1}{3} \ugamma{\mu}\ugamma{\nu}
+\frac{1}{3p^{2}} \ugamma{[\mu}p^{\nu]} \slashed{p}
-\frac{2}{3p^{2}} p^{\mu}p^{\nu} \ ,
%\notag
\\[5pt]
P^{(\frac{1}{2}) \mn}_{11}
&=&\frac{1}{3} \ugamma{\mu}\ugamma{\nu}
-\frac{1}{3p^{2}} \ugamma{[\mu} p^{\nu]} \slashed{p}
-\frac{1}{3p^{2}} p^{\mu}p^{\nu} \ ,
%\notag
\\[5pt]
P^{(\frac{1}{2}) \mn}_{12}
&=& \frac{1}{\sqrt{3}p^{2}} (-p^{\mu}p^{\nu}
+\ugamma{\mu}p^{\nu}\slashed{p}) \ ,
%\notag
\\[5pt]
P^{(\frac{1}{2}) \mn}_{21}
&=& \frac{1}{\sqrt{3}p^{2}} (p^{\mu}p^{\nu}
-\ugamma{\nu}p^{\mu}\slashed{p}) \ ,
%\notag
\\[5pt]
P^{(\frac{1}{2}) \mn}_{22}
&=&\frac{1}{p^{2}} p^{\mu} p^{\nu} \ . \label{eq:Scomponent6} %\notag
\end{eqnarray}
By using the identities shown in Eqs.~(\ref{eqn:spinprojection1})
to (\ref{eqn:spinprojection6}) in Appendix~\ref{app:Deltaprop},
we can immediately write down
\begin{eqnarray}
S_{F}^{0}(p)
&=&P^{(\frac{3}{2})} \frac{-1}{\slashed{p}-m+i \epsilon} P^{(\frac{3}{2})}
\notag \\[5pt]
&&+ P^{(\threehalf \perp)}\left[-\frac{1}{\sqrt{3}m}P^{(\half)}_{12}
-\frac{1}{\sqrt{3}m}P^{(\half) }_{21}
+P^{(\half)}_{22}\frac{2}{3m^{2}}(\slashed{p}+m) P^{(\half)}_{22} \right]
P^{(\threehalf \perp)} \notag \\[5pt]
&\equiv& S_{F}^{0 (\threehalf)} + S_{F}^{0 (\threehalf \perp)} \ .
\label{eqn:Dpropdecomp}
\end{eqnarray}
In principle, the decomposition shown in Eq.~(\ref{eqn:Dpropdecomp})
should be obvious in the beginning, because the Lorentz-invariance
is preserved. However the key is that only the spin-3/2 component
has pole structure, while the spin-1/2 components resemble contact
vertices.

Furthermore, given certain interaction terms, we can carry out the
calculation of the self-energy insertion, as done in
Ref.~\citep{tang98}, for example. Based on the same argument as
given above, the self-energy for renormalization should also be
decomposed into a diagonal form for the spin. The details are as
follows. The self-energy of the $\Delta$ can be defined as
$\Sigma_{\mn} =\Sigma^{\Delta} g_{\mn}+\delta \Sigma_{\mn}$. We see
immediately that $\delta\Sigma_{\mn}$'s indices can only have a
structure like the products of $(\dgamma{\mu}, p_{\mu})
(\dgamma{\nu}, p_{\nu})$. Then we find
\begin{eqnarray}
\Sigma &=&\Sigma^{\Delta} g +\delta \Sigma \\[5pt]
&=& P^{(\frac{3}{2})}\Sigma^{\Delta} g P^{(\frac{3}{2})} +
P^{(\threehalf \perp)} \Sigma  P^{(\threehalf \perp)} \notag \\[5pt]
&&\quad {}+P^{(\frac{3}{2})} (\Sigma^{\Delta} g+\delta \Sigma)
P^{(\threehalf \perp)} + P^{(\threehalf \perp)} (\Sigma^{\Delta}
g+\delta \Sigma) P^{(\frac{3}{2})} \ . %\notag
\end{eqnarray}
So, we can conclude that $\Sigma =P^{(3/2)}\Sigma^{\Delta} P^{(3/2)}
+ P^{(3/2 \perp)} \Sigma  P^{(3/2 \perp)} \equiv \Sigma^{(3/2)} +
\Sigma^{(3/2 \perp)}$. In the proof, we make use of $\left[P^{(3/2)}
\ , \  \Sigma^{\Delta} \right]=0$, $\left[P^{(3/2 \perp)} \ , \
\Sigma^{\Delta} \right]=0$, because the only possible spin
structures of $\Sigma^{\Delta}$ are $\bf{1}$, $\slashed{p}$ and
$\dgammafive$ (parity violation), which commute with the two
projection operators. Then $P^{(3/2)}\Sigma^{\Delta} g P^{(3/2
\perp)}=0$ and $P^{(3/2)\perp}\Sigma^{\Delta} g P^{(3/2)}=0$. Also
we make use of Eqs.~(\ref{eqn:spinprojection2}) and
(\ref{eqn:spinprojection3}), so we get $P^{(3/2)} \delta \Sigma
P^{(3/2 \perp)}=0$ and $P^{(3/2 \perp)} \delta \Sigma P^{(3/2)}=0$.

Based on previous discussions, we can have the following
renormalization of the spin-3/2 propagator:
\begin{eqnarray}
S_{F}&=& (S_{F}^{0 (\threehalf)} + S_{F}^{0 (\threehalf \perp)})
+(S_{F}^{0 (\threehalf)}
+ S_{F}^{0 (\threehalf \perp)}) (\Sigma^{(\frac{3}{2})}
+ \Sigma^{(\frac{3}{2} \perp)})(S_{F}^{0 (\threehalf)}
+ S_{F}^{0 (\threehalf \perp)})+\ldots \notag \\
&=& S_{F}^{0 (\threehalf)}
+ S_{F}^{0 (\threehalf)} \Sigma^{(\frac{3}{2})} S_{F}^{0 (\threehalf)}
+ \ldots \label{eqn:selfenergy1} \\
&&+S_{F}^{0 (\threehalf \perp)}
+S_{F}^{0 (\threehalf \perp)} \Sigma^{(\frac{3}{2} \perp)}
S_{F}^{0 (\threehalf \perp)}+ \ldots \ .  \label{eqn:selfenergy2}
\end{eqnarray}
So the renormalized propagator is decomposed into two different
components: $S_{F}\equiv S_{F}^{(3/2)}+S_{F}^{(3/2 \perp)}$. The
resonant contribution is $S_{F}^{(3/2)}=S_{F}^{0(3/2)}+
S_{F}^{0(3/2)} \Sigma^{(3/2)} S_{F}^{(3/2)}$. The background
contribution is $S_{F}^{(3/2 \perp)}=S_{F}^{0(3/2 \perp)}+
S_{F}^{0(3/2 \perp)} \Sigma^{(3/2 \perp)} S_{F}^{(3/2 \perp)}$. The
renormalization shifts the pole position of the resonant part. For
the nonresonant part, as long as power counting is valid, i.e.,
$O(\Sigma / m) \ll 1$, we are away from any unphysical pole in the
renormalized nonresonant part; $\left[1-O(\Sigma / m)\right]^{-1}$
never vanishes. This also suggests that we will not see the
unphysical pole in the renormalized propagator, when working in the
low-energy perturbative region.
Meanwhile, the argument helps to clarify the redundancy of the
off-shell couplings. We have seen that the self-energy due to these
couplings does not contribute in the renormalization of
$S_{F}^{(3/2)}$. But it indeed changes the nonresonant part.
However, the effect is power expandable. So essentially it is the
same as higher-order contact terms without the $\Delta$. This
justifies the redundancy of these couplings. To ignore them in a way
which does not break chiral symmetry on a term-by-term basis, we can
always associate the $\partial^{\mu}$ with $\pi$ fields so that it
becomes $\ucpartial{\mu}$. \emph{This indicates that those couplings
having $\ucpartial{\mu}$ or $\ugamma{\mu}$ contracted with
$\Delta_{\mu}$ can be ignored without breaking manifest chiral
symmetry.}

A few words on the singularity of $1/p^{2}$ are in order here. [See
Eqs.~(\ref{eq:Scomponent1}) to (\ref{eq:Scomponent6}).] The whole
calculation is only valid in the low-energy limit, and in this limit
we should not find any diagrams with $\Delta$'s that are far ``off
shell". Take pion scattering for example; we assume the pion energy
to be small, and hence $p^{2}$ is always roughly equal to the
incoming nucleon's invariant mass. So the singularity in $1/p^{2}$
should not be a problem in the low-energy theory from a very general
perspective.

\subsection{QHD with \texorpdfstring{$\Delta$}{Delta}}

Consider first $\lag_{\Delta}\ (\hat{\nu}\leqslant
3)$, which is essentially a copy of the corresponding lagrangian for
the nucleon as shown in Eq.~(\ref{eqn:Nlaglowest}):
\begin{eqnarray}
\lag_{\Delta}&=&\frac{-i}{2}\,
          \psibar{\Delta}^{\mkern6mu a}_{\mu}\{\usigma{\mn}\, , \,
          (i\slashed{\ucpartial{}}
          -h_{\rho}\slashed{\rho}-h_{v}\slashed{V}-m
          +h_{s}\phi)\}_{a}^{\mkern3mu b}\, \Delta_{b\nu}
          + \widetilde{h}_{A}\psibar{\Delta}^{\mkern6mu a}_{\mu}
          \slashed{\widetilde{a}}_{a}^{\mkern3mu b} \dgammafive \Delta^{\mu}_{b}
          \notag \\[5pt]
&&{}-\frac{\widetilde{f}_{\rho}h_{\rho}}{4m}\,
          \psibar{\Delta}_{\lambda}\,
          \rho_{\mn}\usigma{\mn}\Delta^{\lambda}
          -\frac{\widetilde{f}_{v}h_{v}}{4m}\,
          \psibar{\Delta}_{\lambda} V_{\mn}\usigma{\mn}\Delta^{\lambda} 
          \notag \\[5pt]
&&{}  -\frac{\widetilde{\kappa}_{\pi}}{m}\,
          \psibar{\Delta}_{\lambda}\widetilde{v}_{\mn}\usigma{\mn}\Delta^{\lambda}
          +\frac{4\widetilde{\beta}_{\pi}}{m}\,
          \psibar{\Delta}_{\lambda}\Delta^{\lambda}\Tr(\widetilde{a}^{\mu}\,
          \widetilde{a}_{\mu}) \ . \label{eqn:Deltalowest}
%\notag
\end{eqnarray}
Here the sub- and superscripts $a, b = (\pm 3/2, \pm 1/2)$, and
the isospin conventions and $T$ matrix have been discussed in
Sec.~\ref{subsec:convention}.

To produce the $N \leftrightarrow \Delta$ transition currents, we
construct the following lagrangians ($\hat{\nu}\leqslant4$):
\begin{eqnarray}
\lag_{\Delta,N,\pi}&=&h_{A}\psibar{\Delta}^{\mkern4mu a
          \mu}\, \Tdagger{iA}{a}\, \widetilde{a}_{i\mu}N_{A} +C.C. \ ,
          \label{eq:transitionhA} %\notag
\end{eqnarray}
\begin{eqnarray}
\lag_{\Delta,\, N,\, \mathrm{background}}&=&
\frac{ic_{1\Delta}}{M}\,
          \psibar{\Delta}^{\mkern4mu a}_{\mu}\dgamma{\nu}\dgammafive\,
          \Tdagger{iA}{a}F_{i}^{(+)\mn}N_{A}
          +\frac{ic_{3\Delta}}{M^{2}}\,
          \psibar{\Delta}^{\mkern4mu a}_{\mu}\, i\dgammafive\, \Tdagger{iA}{a}
          (\dcpartial{\nu}F^{(+)\mn})_{i} N_{A}  \notag \\[5pt]
&&{}+\frac{c_{6\Delta}}{M^{2}}\, \psibar{\Delta}^{\mkern4mu
          a}_{\lambda}
          \dsigma{\mu\nu} \Tdagger{iA}{a}
          (\ucpartial{\lambda}\psibar{F}^{(+)\mn})_{i} N_{A} \notag \\[5pt]
&&{}-\frac{d_{2\Delta}}{M^{2}}\, \psibar{\Delta}^{\mkern4mu
          a}_{\mu}\,
          \Tdagger{iA}{a} (\dcpartial{\nu}F^{(-)\mn})_{i}N_{A}
          -\frac{id_{4\Delta}}{M}\, \psibar{\Delta}^{\mkern4mu a}_{\mu} \dgamma{\nu}\,
          \Tdagger{iA}{a} F_{i}^{(-)\mn}N_{A} \notag \\[5pt]
&&{}-\frac{id_{7\Delta}}{M^{2}}\, \psibar{\Delta}^{\mkern4mu
          a}_{\lambda}
          \dsigma{\mn}\Tdagger{iA}{a}(\ucpartial{\lambda}F^{(-)\mn})_{i}N_{A}
          + C.C. \ , \label{eqn:truelagdeltaNbackground}
\end{eqnarray}
\begin{eqnarray}
\lag_{\Delta,N,\rho}&=&\frac{ic_{1\Delta\rho}}{M}\,
          \psibar{\Delta}^{\mkern4mu a}_{\mu}\,\dgamma{\nu}\dgammafive\,
          \Tdagger{iA}{a}\rho_{i}^{\mn}N_{A}
          +\frac{ic_{3\Delta\rho}}{M^{2}}\, \psibar{\Delta}^{\mkern4mu a}_{\mu}\,
          i\dgammafive\, \Tdagger{iA}{a} (\dcpartial{\nu}\rho^{\mn})_{i} N_{A}
          \notag \\[5pt]
&&{} +\frac{c_{6\Delta\rho}}{M^{2}}\, \psibar{\Delta}^{\mkern4mu
          a}_{\lambda} \dsigma{\mu\nu}\,
          \Tdagger{iA}{a} (\ucpartial{\lambda}\, \psibar{\rho}^{\mkern2mu\mn})_{i} N_{A}
          + C.C. \ .  \label{eqn:truelagdeltaNrho}
\end{eqnarray}
It can be checked that the interaction terms respect all of the
required symmetries. Terms omitted from these lagrangians are either
redundant or are not relevant to the transition interaction
involving $N$ and $\Delta$ (at tree level). The construction of
terms is by means of exhausting all the possibilities. Here we give
an example:
\begin{eqnarray}
\psibar{\Delta}_{\mu}\dgamma{\nu}N \epsilon^{\mn\ab}F^{(+)}_{\ab}
&=&2i\psibar{\Delta}_{\mu} \dgamma{\nu}\dgammafive N F^{(+)\mn}
 +iF^{(+)}_{\ab} \psibar{\Delta}_{\mu} \dgammafive (\ugamma{\mu}\ugamma{\alpha}\ugamma{\beta}-g^{\ab}\ugamma{\mu})N \ .
\end{eqnarray}
The preceding identity indicates that $\psibar{\Delta}^{a}_{\mu}
\dgamma{\nu}\Tdagger{iA}{a} \psibar{F}_{i}^{(+)\mn}N_{A}$ differs
from the $c_{1\Delta}$ coupling in
Eq.~(\ref{eqn:truelagdeltaNbackground}) by off-shell terms, which
can be ignored.

Moreover, the terms in the lagrangian in
Eq.~(\ref{eqn:truelagdeltaNrho}) and the $1/g_{\gamma}$ coupling in
Eq.~(\ref{eqn:lmeson}) are necessary for the realization of
transition form factors using VMD. First, we make the following
definitions:
\begin{eqnarray}
\bra{\Delta, a,p_{\Delta}} V^{i\mu} (A^{i\mu})\ket{N, A,p_{N}} &\equiv&
          \Tdagger{iA}{a}\, \psibar{u}_{\Delta\alpha}(p_{\Delta})\,
          \Gamma_{V (A)}^{\alpha\mu}(q) \, u_{N}(p_{N})
          \ . %\notag \\[5pt]
          \label{eqn:transitioncurrentVAvertex}
\end{eqnarray}
Based on the lagrangians given previously, formulas shown in
Appendix~\ref{app:tildeobjects}, and the definitions of currents in
Eq.~(\ref{eqn:currentdefs}), we find (note that $\sigma_{\mn}
\epsilon^{\mu\nu\alpha\beta} \propto i \sigma^{\alpha\beta}
\dgammafive$)
\begin{eqnarray}
\Gamma_{V}^{\alpha\mu}&=&\frac{2c_{1\Delta}(q^{2})}{M}\,
          (q^{\alpha}\ugamma{\mu}-\slashed{q}g^{\alpha\mu}) \dgammafive
          +\frac{2c_{3\Delta}(q^{2})}{M^{2}}\, (q^{\alpha}q^{\mu}-g^{\alpha\mu}q^{2})
          \dgammafive  \notag \\[5pt]
&&{}-\frac{8c_{6\Delta}(q^{2})}{M^{2}}\,
          q^{\alpha}\usigma{\mn}iq_{\nu}\dgammafive \ ,
          \notag \\[5pt]
c_{i\Delta}(q^{2}) &\equiv& c_{i\Delta}
          +\frac{c_{i\Delta\rho}}{2g_{\gamma}} \frac{q^{2}}{q^{2}-m^{2}_{\rho}}
          \qquad i=1, 3, 6,  \label{eqn:cimd}
\end{eqnarray}
\begin{eqnarray}
\Gamma_{A}^{\alpha\mu}&=&-h_{A} \left(
          g^{\alpha\mu}-\frac{q^{\alpha}q^{\mu}}{q^{2}-m^{2}_{\pi}} \right)
          +\frac{2d_{2\Delta}}{M^{2}}\, (q^{\alpha}q^{\mu}-g^{\alpha\mu}q^{2})
          -\frac{2d_{4\Delta}}{M}\, (q^{\alpha}\ugamma{\mu}-g^{\alpha\mu}\slashed{q})
          \notag \\[5pt]
&& {}-\frac{4d_{7\Delta}}{M^{2}}\, q^{\alpha}\usigma{\mn}iq_{\nu}\ ,
\end{eqnarray}
where $h_{A}$ is from Eq.~(\ref{eq:transitionhA}).
Quite similar to the $c_{i \Delta}(q^{2})$, we can introduce
axial-vector meson [$a_{1}(1260)$] exchange into the axial
transition current, which leads to a structure for the $d_{i
\Delta}(q^{2})$ that is similar to the vector transition current
form factors. There is one subtlety associated with the realization
of $h_{A}(q^{2})$: with our lagrangian, we have the pion-pole
contribution associated only with the $h_{A}$ coupling, and all the
higher-order terms contained in $\delta h_{A}(q^{2})\equiv
h_{A}(q^{2})-h_{A}$ conserve the axial transition current. With the
limited information about manifest chiral-symmetry breaking, we
ignore this subtlety and still use the form similar to the
$c_{1\Delta}(q^{2})$ to parameterize $h_{A}(q^{2})$. The
axial-vector meson couplings $h_{\Delta a_{1}}$ and $d_{i \Delta
a_{1}}$ are the combinations of $g_{a_{1}}$ ($a_{1}$ and isovector
axial-vector external field coupling strength) and the coupling
strength of the $\psibar{\Delta} a_{1} N$ interaction. $m_{a_{1}}$
is the `mass' of the meson. So we have
\begin{eqnarray}
h_{A}(q^{2}) &\equiv& h_{A}+h_{\Delta a_{1}} \,  \frac{q^{2}}{q^{2}-m^{2}_{a_{1}}}\ ,
          \label{eqn:dimd1}\\[5pt]
d_{i\Delta}(q^{2}) &\equiv& d_{i\Delta}
          + d_{i \Delta a_{1}}\, \frac{q^{2}}{q^{2}-m^{2}_{a_{1}}} \
          \qquad i=2, 4, 7.  \label{eqn:dimd2}
\end{eqnarray}

To determine the coefficients in the transition form factors shown
in Eqs.~(\ref{eqn:cimd}), (\ref{eqn:dimd1}), and (\ref{eqn:dimd2}),
we need to compare ours with the conventional ones used in the
literature. In Refs.~\citep{HERNANDEZ07, GRACZYK09} for example, the
definition is
\begin{eqnarray}
\bra{\Delta,\half} j_{cc+}^{\mu} \ket{N,-\half} &\equiv&
          \psibar{u}_{\alpha}(p_{\Delta}) \left\{ \left[
          \frac{C_{3}^{V}}{M}\, (g^{\alpha\mu} \slashed{q}-q^{\alpha}
          \ugamma{\mu}) +\frac{C_{4}^{V}}{M^{2}}\, (q\cdot
          p_{\Delta}\, g^{\alpha\mu}-q^{\alpha} p_{\Delta}^{\mu}) \right. \right. \notag
          \\[3pt]
&&{}+ \left. \frac{C_{5}^{V}}{M^{2}}\, (q\cdot p_{N}\,
          g^{\alpha\mu}-q^{\alpha}
          p_{N}^{\mu}) \right] \dgammafive  \notag
          \\[3pt]
&&{}+\left[ \frac{C_{3}^{A}}{M}\,
          (g^{\alpha\mu}\slashed{q}-q^{\alpha}\ugamma{\mu})
          +\frac{C_{4}^{A}}{M^{2}}\, (q\cdot p_{\Delta}\, g^{\alpha\mu}-q^{\alpha}
          p_{\Delta}^{\mu}) \right. \notag
          \\[3pt]
&&{}+\left. \left. C_{5}^{A}g^{\alpha\mu}+\frac{C_{6}^{A}}{M^{2}}\,
          q^{\mu}q^{\alpha} \right] \right\} u(p_{N}) \ . %\notag
\end{eqnarray}
The basis given above is known to be complete. The determination of
the couplings through comparing our results with the conventional
ones has been given in Ref.~\citep{nunucleon11}. There we find that
our meson dominance form factors are accurate up to $Q^{2} \approx
0.3 \ \mathrm{GeV}^{2}$.
Moreover, CVC and PCAC can be easily checked for the transition
currents. The details can be found in Ref.~\citep{nunucleon11}.

\section{Application}

In this section, we briefly discuss the weak production of pions
from nucleons. We focus only on two properties of the Feynman
diagrams in this problem, including the $G$ parity and the current's
Hermiticity. Then we talk about the production from nuclei, in which
$\Delta$ dynamics is the key component (for both the interaction
mechanism and the final state interaction of the pion). This points
out the importance of understanding the strong interaction,
associated with nuclear structure and $\Delta$ dynamics, in the
study of the electroweak response of nuclei. So it is necessary to
have a framework that includes the two and also provides for
efficient calculations. The details of these subjects are presented
in Refs.~\citep{nunucleon11} and \citep{inunucleus11,cnunucleus11}.

\subsection{Weak production of pions from free nucleons}
\begin{figure}
\centering
\includegraphics[scale=0.5]{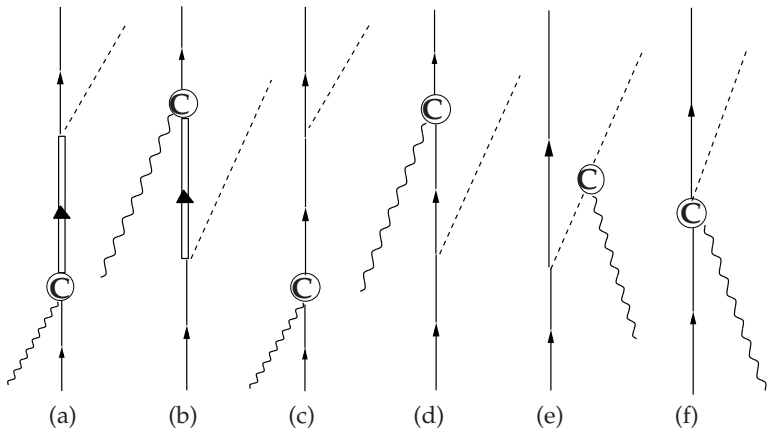}
\put(-270,37){\Large $\bf{C}$} \put(-225,106){\Large $\bf{C}$}
\put(-179,37){\Large $\bf{C}$} \put(-130,104){\Large $\bf{C}$}
\put(-72,78){\Large $\bf{C}$} \put(-33,65){\Large $\bf{C}$}
\put(-270,-10){(a)}
\put(-225,-10){(b)} \put(-180,-10){(c)} \put(-133,-10){(d)}
\put(-87,-10){(e)} \put(-35,-10){(f)} \caption{Feynman diagrams for
pion production. Here, $\bf{C}$ stands for various types of currents
including vector, axial-vector, and baryon currents. Some diagrams
may be zero for some specific type of current. For example, diagrams
(a) and (b) will not contribute for the (isoscalar) baryon current.
Diagram (e) will be zero for the axial-vector current. The pion-pole
contributions to the axial current in diagrams (a) (b) (c) (d)
and (f) are included in the vertex functions of the currents.}
\label{fig:feynmanpionproduction}
\end{figure}
The relevant Feynman diagrams are shown in
Fig.~\ref{fig:feynmanpionproduction} for weak production of pions
due to (anti)neutrino scattering off free nucleons. The `C' in the
figure stands for various currents including the vector current,
axial current, and baryon current, of which both CC and NC are
composed according to Sec.~\ref{subsec:EWsyn}. The details about
these diagrams can be found in \citep{nunucleon11}. Here we begin
with $G$ parity. We use $\bra{N,B,\pi,j} J^{\mu} \ket{N,A}$ to
represent the contribution of diagrams, where `$A$' and `$B$' denote
isospin-1/2 projections. From $G$ parity, we have
\begin{eqnarray}
G A^{i \mu} G^{-1} &=& -A^{i \mu} \ , \notag  \\[5pt]
G V^{i \mu} (J_{B}^{ \mu}) G^{-1} &=& V^{i \mu} (J_{B}^{ \mu}) \ .  \notag
\end{eqnarray}
By applying this to the current's matrix elements, we get
\begin{eqnarray}
\bra{N,B,\pi,j} A^{i\mu} \ket{N,A}
&=&\bra{\psibar{N},B,\pi,j} A^{i\mu} \ket{ \psibar{N},A} \ ,
\label{eqn:crosssymac} 
\end{eqnarray}
\begin{eqnarray}
\bra{N,B,\pi,j} V^{i\mu} (J_{B}^{\mu})\ket{N,A}
&=& -\bra{\psibar{N},B,\pi,j} V^{i\mu} (J_{B}^{\mu}) \ket{ \psibar{N},A}
\ . \label{eqn:crosssymvc}
\end{eqnarray}
Eqs.~(\ref{eqn:crosssymac}) and (\ref{eqn:crosssymvc}) give a
relation between a current's matrix element involving nucleon states
and a matrix element involving antinucleon states. Because of the
isospin symmetry, we can define
\begin{eqnarray}
\bra{N,B,p_{f};\pi,j,k_{\pi}} &A^{i\mu}& \ket{N,A,p_{i}} \notag \\[5pt]
&\equiv&  \delta_{j}^{i} \delta_{B}^{A} \, \psibar{u}(p_{f})
\Gamma^{\mu}_{sym}(p_{f},k_{\pi};p_{i},q) u (p_{i}) \notag \\[5pt]
&&{}+ i \epsilon^{i}_{\;jk} \left(\uhalftau{k}\right)_{B}^{\mkern4mu
A} \psibar{u}(p_{f}) \Gamma^{\mu}_{asym}(p_{f},k_{\pi};p_{i},q)
u(p_{i}) \ . \label{eqn:currentMEdef}
\end{eqnarray}
Vector currents can be decomposed in the same way. From crossing
symmetry, we can see
\begin{eqnarray}
\bra{\psibar{N},B,p_{f};\pi,j,k_{\pi}} &A^{i\mu}& \ket{\psibar{N},A,p_{i}}
\notag \\ [5pt]
&=&  -\delta_{j}^{i} \delta^{A}_{B} \, \psibar{v}(p_{i})
\Gamma^{\mu}_{sym}(-p_{i},k_{\pi};-p_{f},q) v(p_{f}) \notag \\[5pt]
&&{}- i
\epsilon^{i}_{\;jk}\left(-\uhalftau{k}^{T}\right)^{B}_{\mkern4mu A}
\psibar{v}(p_{i})
\Gamma^{\mu}_{asym}(-p_{i},k_{\pi};-p_{f},q) v(p_{f}) \notag \\[5pt]
&=&  \delta_{j}^{i} \delta_{B}^{A} \, \psibar{u}(p_{f})
\left(-\mathcal{C}\Gamma^{T\mu}_{sym}(-p_{i},k_{\pi};-p_{f},q)\mathcal{C}\right) u(p_{i})
\notag \\[5pt]
&&{}- i \epsilon^{i}_{\;jk} \left(\uhalftau{k}\right)_{B}^{\mkern4mu
A} \psibar{u}(p_{f})
\left(-\mathcal{C}\Gamma^{T\mu}_{asym}(-p_{i},k_{\pi};-p_{f},q)
\mathcal{C}\right) u(p_{i})
 \ . \label{eqn:Gtransformedacmatrixelement}
\end{eqnarray}
In Eq.~(\ref{eqn:Gtransformedacmatrixelement}), the
$-\frac{1}{2}\, {\tau^{k}}^{T}$ appears because antiparticles
furnish the complex conjugate representation. (It is equivalent to
the original representation.) $\mathcal{C}$ is the charge conjugate
matrix applied to a Dirac spinor, i.e., $\psi^{C}(x)=\mathcal{C}
(\psibar{\psi}(x))^{T} $. By comparing
Eq.~(\ref{eqn:Gtransformedacmatrixelement}) with
Eq.~(\ref{eqn:crosssymac}), we have the following constraint on the
axial current's matrix element:
\begin{eqnarray}
-\mathcal{C}\Gamma^{T\mu}_{(a)sym}(-p_{i},k_{\pi};-p_{f},q)\mathcal{C}
&=&\underset{(-)}{+}\Gamma^{\mu}_{(a)sym}(p_{f},k_{\pi};p_{i},q) \ . %\notag
\end{eqnarray}
Similarly, we have the following constraint on vector current's
matrix element:
\begin{eqnarray}
-\mathcal{C}\Gamma^{T\mu}_{(a)sym}(-p_{i},k_{\pi};-p_{f},q)\mathcal{C}
&=&\underset{(+)}{-}\Gamma^{\mu}_{(a)sym}(p_{f},k_{\pi};p_{i},q) \ . %\notag
\end{eqnarray}
For the baryon current $\bra{N,B^{\prime},\pi,j} J_{B}^{\mu} \ket{N,A} \equiv
(\frac{1}{2}\, \tau_{j})_{B^{\prime}}^{\mkern4mu A}\, \psibar{u}(p_{f})
\Gamma_{B}^{\mu}(p_{f},k_{\pi};p_{i},q) u(p_{i})$, $G$ parity
indicates
\begin{eqnarray}
-\mathcal{C}\Gamma_{B}^{T\mu}(-p_{i},k_{\pi};-p_{f},q)\mathcal{C}
&=&-\Gamma_{B}^{\mu}(p_{f},k_{\pi};p_{i},q) \ . %\notag
\end{eqnarray}

Now we can see how adding a crossed diagram involving the $\Delta$
is necessary to satisfy $G$ parity. For example, let's talk about
the vector current's matrix element. If we define it for diagrams
(a) and (b) in Fig.~\ref{fig:feynmanpionproduction} as follows:
\begin{eqnarray}
\langle V^{i\mu} \rangle_{a}
&\equiv&  \T{a}{Bj} \Tdagger{iA}{a} \psibar{u}_{f}
\Gamma^{\mu}_{dir}(p_{f},k_{\pi};p_{i},q) u_{i} \ ,
%\notag
\\[5pt]
\langle V^{i\mu} \rangle_{b}
&\equiv&  \T{ai}{B} \Tdagger{A}{ja} \psibar{u}_{f}
\Gamma^{\mu}_{cross}(p_{f},k_{\pi};p_{i},q) u_{i}  \ . %\notag
\end{eqnarray}
Then by using Eq.~(\ref{eqn:isospinprojection}), we get [here we
include only diagram (a) and (b) contributions]
\begin{eqnarray}
\Gamma^{\mu}_{(a)sym} &=&\frac{2}{3}\left(\Gamma^{\mu}_{dir}
\underset{\left(-\right)}{+}  \Gamma^{\mu}_{cross} \right) \ .
\label{eqn:symdircross}
\end{eqnarray}
By calculating the diagrams, it is straightforward to prove that
\begin{eqnarray}
-\mathcal{C}\Gamma^{ T \mu}_{cross}(-p_{i},k_{\pi};-p_{f},q)
\mathcal{C} =-\Gamma^{\mu}_{dir}(p_{f},k_{\pi};p_{i},q) \ .  %\notag
\end{eqnarray}
This equation justifies the $G$ parity of the vector current's
matrix elements. Other currents' matrix elements can be justified in
a similar way.

Now we discuss the Hermiticity of the current. Let's consider
$\bra{N,\pi \, out} J^{\mu} \ket{N, \, in} $:
\begin{eqnarray}
\bra{N,p_{f},\pi,k_{\pi}, out} J^{\mu} \ket{N,p_{i}, in}^{\ast}
 &=&\bra{N,p_{i}, in} J^{\dagger \mu} \ket {N,p_{f},\pi,k_{\pi}, out}
 \notag \\[5pt]
&\neq&
\bra{N,p_{i},out} J^{\dagger \mu} \ket {N,p_{f},\pi, k_{\pi}, in} \ . %\notag
\end{eqnarray}
But how do we generally understand $\bra{i,in} O \ket {f, out}$?
Naively, we would have the following:
\begin{eqnarray}
\bra{i,in} O \ket {f, out}
&=& \bra{i} U(-\infty,0) U(0,t)o(t)U(t,0) U(0,+\infty) \ket {f} \notag \\[5pt]
&=&\bra{i} \psibar{T}o(t) \exp [i\int dt H_{I}(t)] \ket {f} \ .
\end{eqnarray}
Here $O$ and $o(t)$ are the operators in the Heisenberg and
interaction pictures. $\psibar{T}$ is another type of time ordering:
$\psibar{T} H_{I}(t_{1})H_{I}(t_{2})=\theta(t_{2}-t_{1})
H_{I}(t_{1})H_{I}(t_{2})+\theta(t_{1}-t_{2})
H_{I}(t_{2})H_{I}(t_{1})$.  It is easy to realize that in momentum
space, if we mirror the pole of the $T$ defined Green's function,
and apply $(-)$ to the overall Green's function, we get the
$\psibar{T}$ defined Green's function. Second, each interaction
vertex in the $\bra{i,in} O \ket {f, out}$ calculation differs from
that of $\bra{i,out} O \ket {f, in}$ by a $(-)$ sign. Third, since
now all the poles are in the first and third quadrants in the
complex momentum plane, the corresponding loop integration differs
from the normal loop integration by a $(-)$ sign! So, without a
rigorous proof, we have that after calculating $\bra{i , out} O \ket
{f , in}$, if we mirror all the poles relative to the real axis for
the propagator and apply a phase $(-)^{(V-V_{o})+I+L}=(-)^{V_{o}-1}$
to it, then we get the corresponding $\bra{i , in} O \ket {f ,
out}$. Here $V,V_{o},I$, and $L$ are the number of vertices in the
graph, vertices in the operator $O$, internal lines, and loops. For
the current operator $J^{\mu}$, $V_{o}=1$ and hence the phase is
$(+)$.

Now let's proceed to see the consequence of the Hermiticity of $J^{
\mu}(x=0)$, i.e., $J^{i \mu \dagger}=J_{i}^{\mu}$:
\begin{eqnarray}
\bra{N,B,p_{f},\pi,j,k_{\pi}, out} &\! J^{i \mu}\! &
       \ket{N,A,p_{i}, in}^{\ast} \notag \\[5pt]
&=&\bra{N,A,p_{i}, in} J_{i}^{\mu} \ket {N,B,p_{f},\pi,j,k_{\pi}, out}
\notag \\[5pt]
&=& \bra{N,A,p_{i},out} J_{i}^{\mu}
    \ket {N,B,p_{f},\pi,j, k_{\pi}, in} \vert_{pm} \notag \\[5pt]
&=&  \delta_{i i^{\prime}} \delta^{j j^{\prime}}
     \bra{N,A,p_{i},\pi,j^{\prime},-k_{\pi}, out} J^{ i^{\prime}\mu}
     \ket {N,B,p_{f}, in}\vert_{pm} \ .
\label{eqn:hermicityconstraint}
\end{eqnarray}
Here $\vert_{pm}$ indicates poles are mirrored with respect to the
real axis. In the following, we decompose the general current matrix
element into symmetric and antisymmetric parts, as we did in in
Eq.~(\ref{eqn:currentMEdef}):
\begin{eqnarray}
&&\bra{N,B,p_{f},\pi,j,k_{\pi}, out} J^{i \mu} \ket{N,A,p_{i}, in}^{\ast}
  \notag \\[5pt]
&=&\delta^{j}_{i} \delta^{B}_{A}\, \psibar{u}(p_{i})
   \psibar{\Gamma}^{\mu}_{sym}(p_{f},k_{\pi};p_{i},q) u (p_{f})
   - i \epsilon_{i}^{\;jk} \left(\dhalftau{k}\right)^{\mkern4mu B}_{A} \psibar{u}(p_{i})
    \psibar{\Gamma}^{\mu}_{asym}(p_{f},k_{\pi};p_{i},q) u(p_{f})
   \ .   \label{eqn:complexconjugatecurrent}
\end{eqnarray}
Here, $\psibar{\Gamma}=\ugamma{0}\Gamma^{\dagger}\ugamma{0}$.
Meanwhile, Eq.~(\ref{eqn:hermicityconstraint}) can be rewritten as
\begin{eqnarray}  
\bra{N,A,p_{i},\pi,j^{\prime},-k_{\pi}, out} &\! J^{ i^{\prime}\mu}
\!&
\ket {N,B,p_{f}, in} \vert_{pm} \, \delta_{i i^{\prime}} \delta^{j j^{\prime}}  \notag \\[5pt]
&=&\delta_{i i^{\prime}} \delta^{j j^{\prime}}
\bigg[\delta_{j^{\prime}}^{i^{\prime}} \delta_{A}^{B}\,
\psibar{u}(p_{i})
\Gamma^{\mu}_{sym}(p_{i},-k_{\pi};p_{f},-q) u (p_{f})  \notag \\[5pt]
&&{}+ i \epsilon^{i^{\prime}}_{\;j^{\prime}k}
\left(\uhalftau{k}\right)_{A}^{\mkern4mu B}
      \psibar{u}(p_{i}) \Gamma^{\mu}_{asym}(p_{i},-k_{\pi};p_{f},-q) u(p_{f})
  \bigg]_{pm} \notag \\[5pt]
&=&\delta^{j}_{i} \delta_{A}^{B}\, \psibar{u}(p_{i})
\Gamma^{\mu}_{sym}(p_{i},-k_{\pi};p_{f},-q) u (p_{f}) \vert_{pm} \notag \\[5pt]
&&{}+ i \epsilon_{i}^{\;jk} \left(\dhalftau{k}\right)_{A}^{\mkern4mu
B} \psibar{u}(p_{i}) \Gamma^{\mu}_{asym}(p_{i},-k_{\pi};p_{f},-q)
u(p_{f})\vert_{pm} \ . \label{eqn:pm1}
\end{eqnarray}
If we compare Eq.~(\ref{eqn:complexconjugatecurrent}) with
Eq.~(\ref{eqn:pm1}), we see the Hermiticity constraint is
\begin{eqnarray}
\ugamma{0} [\Gamma^{ \mu}_{(a)sym}(p_{f},k_{\pi};p_{i},q)]^{\dagger}
\ugamma{0}
&=&\underset{(-)}{+}\Gamma^{\mu}_{(a)sym}(p_{i},-k_{\pi};p_{f},-q)\vert_{pm}
\ . \label{eqn:hermicityconstraint1}
\end{eqnarray}

Now let's focus on the constraint on diagrams (a) and (b) in
Fig.~\ref{fig:feynmanpionproduction}. We can check by calculating
diagrams:
\begin{eqnarray}
\psibar{\Gamma}^{\mu}_{dir}(p_{f},k_{\pi};p_{i},q)
&=&\Gamma^{\mu}_{cross}(p_{i},-k_{\pi};p_{f},-q)\vert_{pm} \ .  %\notag
\end{eqnarray}
We can choose kinematics where no poles and cuts arise, i.e., there
is no phase shift, and then test the constraint without $|_{pm}$.
The preceding observation, with Eq.~(\ref{eqn:symdircross}) taken
into account, leads to the satisfaction of the constraint in
Eq.~(\ref{eqn:hermicityconstraint1}). The Hermiticity of the baryon
current can be studied in a similar way, and hence is not shown
explicitly here. Moreover, it is interesting to see that the
higher-order contact terms satisfy the requirements due to $G$
parity and Hermiticity on a term-by-term basis.

\subsection{Weak production of pions from nuclei, \texorpdfstring{$\Delta$}{Delta} dynamics}
With the development of neutrino-oscillation experiments, precise
knowledge about the neutrino (antineutrino)-nuclei scattering cross
sections is needed for the understanding of the experiments'
background. Take MiniBooNE \citep{MiniBN2010, MiniBN2010anti}, for
example; the median energy of the neutrino (antineutrino) beam is
around 0.6 (0.5) GeV, and the high-energy tail extends up to 2 GeV.
In this regime, the $\Delta$ is the most important resonance for the
interaction mechanism, except in the \emph{very} low-energy region.
Therefore, to understand pion production, we need to study $\Delta$
dynamics in the nucleus. This subject has been extensively discussed
in the nonrelativistic framework \citep{oset87, hirata79,
horikawa80}, and it has also been initiated in the relativistic
framework in \citep{wehrberger89, wehrberger90, wehrberger92,
wehrberger93}. It is shown that the $\Delta$ width increases in the
normal nuclear medium, since new decay channels are opened, like
$\Delta N \rightarrow N N$, for example. The real part of the
$\Delta$'s self-energy has also been studied. From the lagrangian in
Eq.~(\ref{eqn:Deltalowest}), we can see that the two parameters
$h_{s}$ and $h_{v}$ in the lagrangian are
important.\footnote{$h_{\rho}$ should not play an important role in
normal nuclei with small asymmetry.} However, the information in
\citep{wehrberger89, wehrberger93, boguta81, kosov98} is still
limited. In \citep{inunucleus11,cnunucleus11}, we have realized that
these $\Delta$-meson couplings are responsible for the $\Delta$'s
spin-orbit coupling in the nucleus, and based on this we provide
some information about the couplings from this new perspective.

Meanwhile, the $\Delta$ dynamics is also strongly correlated with the
pion dynamics in the nuclear medium, and hence is important for
understanding the pion's final state interactions, especially in the
energy regime of these neutrino-oscillation experiments.

\section{Summary}
In this work, we have studied EW interactions in QHD EFT. First, we
discuss the EW interactions at the quark level. Then we include EW
interactions in QHD EFT by using the background-field technique. The
completed QHD EFT has a nonlinear realization of $SU(2)_L \otimes
SU(2)_R \otimes U(1)_B$ (chiral symmetry and baryon number
conservation), as well as realizations of other symmetries including
Lorentz-invariance, $C$, $P$, and $T$. Meanwhile, as we know, chiral
symmetry is manifestly broken due to the nonzero quark masses; the $P$
and $C$ symmetries are also broken because of weak interactions. All
these breaking patterns are parameterized in a general way in the
EFT. Moreover, we have included the $\Delta$ resonance as manifest
degrees of freedom in our QHD EFT. This enables us to discuss
physics at the kinematics where the resonance becomes important. As
a result, the effective theory uses hadronic degrees of freedom,
satisfies the constraints due to QCD (symmetries and their breaking
pattern), and is calibrated to strong-interaction phenomena. (The EW
interaction of \emph{individual hadrons}, like the transition currents
discussed in this work, need to be parameterized.) So this effective
field theory satisfies the three listed points laid out in the
Introduction.

The technical issues that arise when introducing the $\Delta$ in the
EFT need to be emphasized here. It has been proven that the general
EFT with conventional interactions has no redundant degrees of
freedom \citep{Krebs09}. (Unphysical degrees of freedom have been
considered in the canonical quantization scheme as the reason for
pathologies in field theory with high-spin fields.) However, the
proof rests on the work of \citep{Pascalutsa98}, which claims that
gauge invariance could eliminate the redundant degrees of freedom.
Here, we have provided another perturbative argument about this
issue, which indicates that as long as we work in the low-energy and
weak-field limit, the unphysical degrees of freedom do not show up.
This condition is satisfied in the EFT. Throughout the argument, we
do not need to make use of the gauge-invariance requirement. And in
this way, we can easily see the redundancy of \emph{off-shell}
interactions, which has also been rigorously addressed in
\citep{Krebs10}. Moreover, the argument can be easily generalized to
other high-spin fields.

To appreciate the importance of the symmetries realized in QHD EFT,
we have discussed the currents' matrix elements in pion production
from nucleons. The calculation and results are detailed in
\citep{nunucleon11}. Here, we first briefly mention the consequence
of chiral symmetry (and its breaking), i.e., CVC and PCAC. These two
principles provide important constraints on the EW interactions at
the hadronic level. The $G$ parity is then studied for pion
production. This provides another constraint on the analytical
structure of matrix elements. Meanwhile, it also points out the
importance of including cross diagrams involving the $\Delta$. When
combining the $\Delta$'s contribution in the $s$ and $u$ channels,
the full result respects $G$ parity. Moreover, the constraint due to
the Hermiticity of current operators is explored. It is important to
notice that other contact terms respect all these constraints. So,
it is necessary to have a theoretical framework that satisfies these
constraints. The QHD EFT, with symmetries included, clearly provides
such a framework.

However, the calibration of a model on the hadronic level does not
guarantee its success at the nuclear level. To study EW interactions
in nuclei, we clearly have to understand how the nucleons are bound
together to form nuclei. QHD has been applied extensively to this
kind of problem \citep{Serot86,Serot97}, and the recently developed
chiral QHD EFT has also been tested in the nuclear many-body problem
\citep{Furnstahl9798}. The mean-field approximation is understood in
terms of density functional theory \citep{Kohn99}, and hence the
theory calibrated to nuclear properties includes many-body
correlations beyond the Hartree approximation. Moreover, the power
counting of diagrams in terms of $O(k/M)$ ($k$ can be the Fermi
momentum, mean-field strength, or other dimensional quantities) in
the many-body calculations has also been studied in this framework
with the justification that fitted parameters are \emph{natural}
\citep{Twoloops1, Twoloops2}. This enables us to discuss the EW
interactions order-by-order in the nuclear many-body system using
QHD EFT.

As mentioned before, we have initiated the study of weak production
of pions due to neutrino and antineutrino scattering off nuclei in
this framework \citep{nunucleon11,inunucleus11,cnunucleus11}.
Moreover, we also studied the production of photons, in which the
conservation of the EM current is clearly crucial. The discussion of
power counting has been presented in these references. Furthermore,
we should also anticipate the importance of $\Delta$ dynamics
modified in nuclei. It has been studied in the nonrelativistic
framework, but just started in the relativistic framework. The study
indicates that the $\Delta$ decay width increases at normal nuclear
density because the reduced pion-decay phase space is more than
compensated by the opening of other decay channels. But a detailed
discussion on this is still needed. The real part of the $\Delta$
self-energy is still unclear. As we pointed out, the $h_{s}$ and
$h_{v}$ couplings in Eq.~(\ref{eqn:Deltalowest}) play important
roles, but there are still limited constraints on them. (Some
constraints have been gained from an equation of state perspective,
and others come from electron scattering.) As we realized in
\citep{inunucleus11,cnunucleus11}, the phenomenologically fitted
spin-orbit coupling of the $\Delta$ in the nucleus may shed some
light on this issue. Clearly, more efforts are needed to study
$\Delta$ dynamics, which in the meantime is closely related to pion
dynamics in the nuclear many-body system.

\section*{Acknowledgements}
This work was supported in part by the US Department of Energy 
under Contract No. DE--FG02--87ER40365.
\appendix

\section{Isospin indices, \texorpdfstring{$T$}{T} matrices}\label{app:indices}
Suppose $\vec{t}$ are the generators of some (ir)reducible
representation of $SU(2)$; then it is easy to prove that
($\underline{\widetilde\delta}\equiv-\underline{e}^{-i\pi
t^{y}}$)
\begin{eqnarray}
(-\vec{t}^{\mkern4muT})^{\;i}_{\;\;j}&=&\widetilde\delta^{ik} \, \vec{t}_{k}^{\;\;l} \, \widetilde\delta_{lj}  \equiv  \vec{t}^{\;i}_{\;\;j}  \quad
\ , \text{i.e.} \ , \quad
-\underline{\vec{t}}^{\mkern6muT}=\underline{\widetilde\delta} \ \underline{\vec{t}} \
\underline{\widetilde\delta}^{\mkern4mu
-1} \ .%\notag
\end{eqnarray}
Here the superscript $T$ denotes transpose. This equation justifies
the use of $\widetilde{\delta}$ as a metric linking the
representation and the equivalent complex-conjugate representation. One easily
finds for $\mathcal{D}^{(3/2)}$, $\mathcal{D}^{(1)}$, and
$\mathcal{D}^{(1/2)}$,
\begin{eqnarray}
\widetilde\delta^{ab}=
                     \begin{pmatrix}
                     0&0&0&1 \\
                     0&0&-1&0 \\
                     0&1&0&0\\
                     -1&0&0&0
                     \end{pmatrix} \ ,&\qquad&
\widetilde\delta_{ab}=
                     \begin{pmatrix}
                     0&0&0&-1 \\
                     0&0&1&0 \\
                     0&-1&0&0\\
                     1&0&0&0
                     \end{pmatrix} \ ,
\end{eqnarray}
\begin{eqnarray}
\widetilde\delta^{\mkern3mu ij}=\begin{pmatrix}
                     0&0&-1 \\
                     0&1&0 \\
                    -1&0&0
                      \end{pmatrix} \ , &\qquad&
\widetilde\delta_{\mkern3mu ij}=\begin{pmatrix}
                     0&0&-1 \\
                     0&1&0 \\
                    -1&0&0
                      \end{pmatrix}    \ ,  \\[5pt]
\widetilde\delta^{AB}=
                     \begin{pmatrix}
                     0&1 \\
                     -1&0
                     \end{pmatrix} \ , &\qquad&
\widetilde\delta_{AB}=
                     \begin{pmatrix}
                     0&-1 \\
                     1&0
                     \end{pmatrix} \ .  
\end{eqnarray}

We now turn to the $T$ matrices. As discussed in
Sec.~\ref{subsec:convention},
\begin{eqnarray}
T^{\dagger \,\,\,iA}_{a}=\langle\frac{3}{2}; a \vert 1,\frac{1}{2};i,A
\rangle \ , \qquad
T^{a}_{\,\,iA}=\langle  1, \frac{1}{2};i,A \vert \frac{3}{2};
a\rangle \ . %\notag
\end{eqnarray}
It is easy to prove the following relations (here $\tau^{i}$ is a
Pauli matrix):
\begin{eqnarray}
\tau^{i} \, \tau_{j} &=& \widetilde\delta^{\mkern3mu i}_{\mkern3mu
          j} + i\, \widetilde\epsilon^{\mkern5mu i}_{\;jk} \tau^{k} \ ,
          \\[5pt]
\left(P_{i}^{j}\right)_{A}^{\; B} &\equiv& T^{a}_{\;iA}\,
          T^{\dagger \;jB}_{a} =\widetilde\delta_{\mkern3mu i}^{j}
          \,\widetilde\delta_{A}^{\;B}-\frac{1}{3}(\tau_{i}\tau^{j})_{A}^{\;B} \ ,
          \label{eqn:isospinprojection} \\[5pt]
T^{\dagger \;iA}_{a}\;
          T^{b}_{\;iA}&=&\widetilde\delta_{a}^{\mkern3mu b} \ . %\notag
\end{eqnarray}
Here $P_{i}^{j}$ is a projection operator that projects
$\mathcal{H}^{(1/2)} \otimes \mathcal{H}^{(1)}$ onto
$\mathcal{H}^{(3/2)}$.

A few words about $\widetilde\epsilon^{\mkern5mu i}_{\;jk}$ are in
order here. We have the following transformations of pion fields:
\begin{eqnarray}
\pi^{i}  &=& \pi^{I} u^{\mkern3mu i}_{I}   \qquad \qquad
\text{here, $i=+1,0,-1$}\ ; \quad \text{ $I=x,y,z$ } \ , \notag \\[5pt]
\begin{pmatrix}
\pi^{+1} , \
\pi^{0} , \
\pi^{-1}
\end{pmatrix} &=&
\begin{pmatrix}
\pi^{x} , \
\pi^{y} , \
\pi^{z}
\end{pmatrix}
\begin{pmatrix}
\displaystyle{\frac{-1}{\sqrt{2}}} &0 & \displaystyle{\frac{1}{\sqrt{2}}}
\\[10pt]
\displaystyle{\frac{-i}{\sqrt{2}}} &0 & \displaystyle{\frac{-i}{\sqrt{2}}}
\\[10pt]
0 &1& 0
\end{pmatrix} \ . \label{eqn:piondef}
\end{eqnarray}
Under such transformations,
\begin{eqnarray}
\widetilde\epsilon^{\mkern5mu ijk}&\equiv& u^{i}_{I}\, u^{j}_{J}\, u^{k}_{K}
             \, \epsilon^{IJK}=\det(\underline{u^{i}_{I}})
             \epsilon^{ijk}=-i\, \epsilon^{ijk} \notag  \\
\Longrightarrow \quad\widetilde\epsilon^{\mkern5mu ijk}&=&
 \begin{cases}
  -i,            & \text{if $ijk=+1,0,-1$} \ ; \\
  -i \,\delta_{\mathcal{P}}, & \text{if $ijk=\mathcal{P}(+1,0,-1)$} \ .
 \end{cases}  %\notag
\end{eqnarray}
Here $\delta_{\mathcal{P}}$ is the phase related with the
$\mathcal{P}$ permutation. It is $+$ ($-$) with an even (odd) number
of permutations. To simplify the notation, we will ignore the
\emph{tilde} on $\widetilde{\delta}$ and $\widetilde\epsilon\,$ in
other places.

\section{Expansion of tilde objects} \label{app:tildeobjects}
Here we show some details about $\widetilde{v}_{\mu}$,
$\widetilde{a}_{\mu}$, $F^{(\pm)}_{\mn}$ and others, which
are needed for understanding electroweak interactions in
QHD EFT. The pion-decay constant is $f_{\pi}\approx 93$ MeV.
\begin{eqnarray}
\Tr(\uhalftau{i}[U\ , \partial^{\mu}U^{\dagger}])&\approx&
        2i\epsilon^{ijk}\,\frac{\pi_{j}}{f_{\pi}}\,
        \frac{\partial_{\mu}\pi_{k}}{f_{\pi}}\ ,  \label{eqn:upartialu+c}\\
\Tr(\uhalftau{i}\{U\ , \partial^{\mu}U^{\dagger}\})&\approx& -2i\,
        \frac{\partial_{\mu}\pi^{i}}{f_{\pi}}\ , \label{eqn:upartialu+ac} 
\end{eqnarray}
\begin{eqnarray}
\xi^{\dagger} \uhalftau{i}\xi + \xi\uhalftau{i}\xi^{\dagger}&\approx&\tau^{i}\ , \\
\xi^{\dagger} \uhalftau{i}\xi - \xi\uhalftau{i}\xi^{\dagger}&\approx&
        -\epsilon^{ijk}\, \frac{\pi_{j}}{f_{\pi}}\, \tau_{k}\ , 
\end{eqnarray}
\begin{eqnarray}
\widetilde{v}_{\mu}&\approx& \frac{1}{2f^{2}_{\pi}}\,
        \epsilon^{ijk}\pi_{j}\partial_{\mu}\pi_{k}\dhalftau{i}
        -\vbg_{i\mu}\uhalftau{i}-\epsilon^{ijk}\, \frac{\pi_{j}}{f_{\pi}}\, \dhalftau{k}\, \abg_{i\mu}\ ,
        \label{eqn:vmu}\\
\widetilde{a}_{\mu}&\approx& \frac{1}{f_{\pi}}\,
        \partial_{\mu}\pi^{i}\, \dhalftau{i}
        +\abg_{i\mu}\uhalftau{i}+\epsilon^{ijk}\, \frac{\pi_{j}}{f_{\pi}}\, \dhalftau{k}\vbg_{i\mu}\ ,
        \label{eqn:amu}  \\ 
\widetilde{v}_{\mn}&\approx&\frac{1}{f^{2}_{\pi}}\,
        \epsilon^{ijk}\partial_{\mu}\pi_{j}\partial_{\nu}\pi_{k}
        \, \dhalftau{i} \notag \\[5pt]
&&{}-\left( i \left[\frac{1}{f_{\pi}}\, \partial_{\mu}\pi^{i}\,\dhalftau{i}\, , \,
        \abg_{\nu}+\epsilon^{ijk}\, \frac{\pi_{j}}{f_{\pi}}\, \dhalftau{k}\, \vbg_{i\nu} \right]
        -(\mu \leftrightarrow \nu) \right)  \notag \\[5pt]
&&{}+ \text{background\ interference\ terms,} \label{eqn:vmn}\\[5pt]
\rho_{\mn}&=&\partial_{[\mu}\rho_{\nu
        ]}+i\overline{g}_{\rho}[\rho_{\mu}\ , \ \rho_{\nu}]
        + i ([\widetilde{v}_{\mu}\ , \ \rho_{\nu}] - \mu \leftrightarrow \nu)\ ,
        \label{eqn:rhomn} \\[5pt]
f_{L\mn}+f_{R\mn}&=& 2\partial_{[\mu}\vbg_{\nu]} -2i[\vbg_{\mu}\ , \ \vbg_{\nu}]
        -2i[\abg_{\mu}\ , \ \abg_{\nu}]\ , \\[5pt]
f_{L\mn}-f_{R\mn}&=& -2\partial_{[\mu}\abg_{\nu]} +2i[\vbg_{\mu}\ , \ \abg_{\nu}]
        +2i[\abg_{\mu}\ , \ \vbg_{\nu}]\ , \\[5pt]
F^{(+)}_{\mn}
&\approx& 2\partial_{[\mu}\vbg_{\nu]} +2 \epsilon^{ijk}\,
        \frac{\pi_{j}}{f_{\pi}}\,\dhalftau{k}\,
        \partial_{[\mu}\abg_{i\nu]} + \text{background\ interference,} \label{eqn:F+mn}\\
F^{(-)}_{\mn}
&\approx& -2\partial_{[\mu}\abg_{\nu]} -2
        \epsilon^{ijk}\, \frac{\pi_{j}}{f_{\pi}}\, \dhalftau{k}\, \partial_{[\mu}\vbg_{i\nu]}
        + \text{background\ interference.} \label{eqn:F-mn}
\end{eqnarray}

\section{Properties of projection operators in the spin-3/2 propagator} \label{app:Deltaprop}
We have properties about these spin projectors:
\begin{eqnarray}
(P^{(I) }_{ij})^{\mn} (P^{(J)}_{kl })_{\nu \lambda}
&=&\delta_{I J} \delta_{jk} (P^{(I)}_{il})^{\mu}_{\lambda}
\ , \label{eqn:spinprojection1} \\[5pt]
\ugamma{\mu} P^{\left(\frac{3}{2}\right)}_{\mn}
&=&P^{\left(\frac{3}{2}\right)}_{\mn} \ugamma{\nu} =0 \ ,
\label{eqn:spinprojection2} \\[5pt]
p^{\mu}P^{\left(\frac{3}{2}\right)}_{\mn}
&=&P^{\left(\frac{3}{2}\right)}_{\mn} p^{\nu} =0 \ .
\label{eqn:spinprojection3}
\end{eqnarray}
Based on the above identities, we can prove that
\begin{eqnarray}
P^{(\threehalf)}+P^{(\half)}_{11}+P^{(\half)}_{22}&=& \bf{1}  \ ,
\label{eqn:spinprojection4} \\[5pt]
 P^{(\half)}_{11}+P^{(\half)}_{22}& \equiv& P^{(\threehalf \perp) }
 \ , \label{eqn:spinprojection5}  \\[5pt]
\left[P^{(\threehalf)}  \ , \  \slashed{p}\right]=\left[P^{(\half)}_{11} \ , \ \slashed{p}\right]&=&\left[P^{(\half)}_{22}  \ , \  \slashed{p}\right]=0 \ .
\label{eqn:spinprojection6} 
\end{eqnarray}

%%%%%%%%%%%%%%%%%%%%%%%%%%%%%%%%

\end{document}